\documentclass[useAMS,usenatbib,a4paper]{mn2e}

\usepackage{multirow}
\usepackage{graphicx}
\usepackage{subfigure}
\usepackage{amsmath}
\usepackage{amsfonts}

\DeclareMathOperator{\divergence}{div}
\DeclareMathOperator{\gradient}{grad}

\title[Multidimensional modelling of Cepheid-like variables II:\, \\
Analysis of a Cepheid model]{
Multidimensional realistic modelling of Cepheid-like variables II:\, \\
Analysis of a Cepheid model}
\author[E. Mundprecht, H.J. Muthsam and F. Kupka]{Eva Mundprecht$^{1}\thanks{E-mail:
eva.mundprecht@univie.ac.at}$, Herbert J. Muthsam$^{1}\thanks{E-mail:
herbert.muthsam@univie.ac.at}$ and Friedrich Kupka$^{1}$\\
$^{1}$University of Vienna, Faculty of Mathematics, Oskar-Morgenstern-Platz 1, A-1090 Wien, Austria}

\begin{document}

\date{Accepted ..... Received ......; in original form...}

\pagerange{\pageref{firstpage}--\pageref{lastpage}} \pubyear{xxxx}

\maketitle

\label{firstpage}

\begin{abstract}
Non-local, time-dependent convection models have been used to explain the location of double-mode pulsations in Cepheids in the HR diagram as well as the existence and location of the red edge of the instability strip. These properties are highly sensitive to model parameters. We use 2D radiation hydrodynamical simulations with realistic microphysics and grey radiative-transfer to model a short period Cepheid.\\
The simulations show that the strength of the convection zone varies significantly over the pulsation period and exhibits a phase shift relative to the variations in radius. We evaluate the convective flux and the work integral as predicted by the most common convection models. It turns out that over one pulsation cycle the model parameter $\alpha_{\rm c}$, has to be varied by up to a factor of beyond 2 to match the convective flux obtained from the simulations. To bring convective fluxes integrated over the He~{\sc ii} convection zone and the overshoot zone below into agreement, this parameter has to be varied by a factor of up to $\sim 7.5$ (Kuhfu{\ss}).\\
We then present results on the energetics of the convection and overshoot zone by radially symmetric and fluctuating quantities.
To successfully model this scenario by a static, one dimensional or even by a simple time-dependent model appears extremely challenging. We conclude that significant improvements are needed to make predictions based on 1D models more robust and to improve the reliability of conclusions on the convection-pulsation coupling drawn from them. Multidimensional simulations can provide guidelines for developing descriptions of convection then applied in traditional 1D modelling.
\end{abstract}


\section{Introduction}
In the first paper of this series, \citet{mund13} (paper I henceforth), we have described extensions of the ANTARES code (\citealt{muth_na10}) which enable the investigation of the pulsation-convection interaction by numerically solving the equations of radiation-hydrodynamics with realistic microphysics and opacities. 
We now turn to the description and analysis of a 2D nonlinear Cepheid model. While the first paper centered around technical issues, the present one concentrates on the physical properties of the convection zone caused by the ionisation of He~{\sc ii} which also hosts much of the pulsational driving exhibited by the $\kappa$-mechanism. To motivate this work we first discuss here
a few items related to the physics of Cepheid pulsations which explains the interest in the modelling of the pulsation-convection
interaction. Rewards from such work are expected in the course of time, both in the immediate and in the later future.

The success of the effort put into modelling the pulsation-convection interaction has, to a large degree, been measured by the
accuracy up to which properties of \textit{individual Cepheids} can be reproduced (light curve, in particular also special features
such as humps and bumps, etc.). There is now a large body of 1D simulations which are utilized in various astrophysical contexts.
Some problems and trends in Cepheid research have been described in two reviews, \citet{buch_rev97} and \citet{buch_rev09}. The
status and progress in research on the related RR Lyr stars has recently been reviewed by \citet{marc_rev09}. It is remarkable
that both authors stress open problems in the treatment of pulsation-convection interaction. Instead of repeating the problems
discussed in those reviews we want to shortly describe a few problem areas which are either different from those discussed there
or are considered here from a different point of view, and where better insights to the pulsation-convection interaction can
expected to be crucial for progress.

Let us, first of all, point out that in the usual 1D simulations for those stars convection is modelled by simplified non-local, time-dependent models, specifically adapted to an environment varying due to the radial pulsations. Generally, such models need some equation for the turbulent kinetic energy, and due to the variable environment at least a time evolution equation for the turbulent kinetic energy is necessary. This is in contrast to the possibly static equation for the turbulent energy which is applicable in stars which do not exhibit pulsation. See the discussion in \citet{buch_koll_00} where the most widely used models are described. 

These models require, however, to know a number of parameters, the so-called $\alpha$'s, which are unknown a priori. 
Typically, these parameters are calibrated so that models reproduce properties which those objects exhibit, for example,
the existence and the location of the red edge of the pulsation instability strip. Sometimes they are simply guessed.
Since the number of these parameters is, in addition, considerable, and even the form of the added equations or closure approximations 
may well be ambiguous concern must be expressed as to the reliability of this approach.

As a consequence, these models must be tested and, if feasible, the parameters calibrated in different, independent ways. 
Presently, it is not known to what extent the results obtained from conventional models are affected, at least in the case of 
cooler pulsating stars where effects due to convection are largest. It is therefore of interest to consider a specific problem where
what had once been considered progress has again been challenged just on grounds of inappropriate modelling of convection:
the (1D) simulation of double mode Cepheids. 

In parameter regions where double mode behaviour is observed, nonlinear, radially symmetric models for pulsating stars
have failed to reproduce such a behaviour for a long time. It has therefore been considered to be an important breakthrough
when, ultimately, models did exhibit double mode pulsation indeed. These results have, however, been 
criticized by \citet{smolec_mo08}, where also references to other papers are given, on grounds of inappropriate modelling 
of convection, in particular because of much too large overshoot. Indeed, the question of how to explain double mode pulsation
in Cepheids is now considered open again.

Quite generally, these closure parameters of the convection models are known to depend on effective temperature,
chemical composition, etc., insofar as they are can be determined at all with some reasonable reliability. This holds already true for the "classical" mixing-length parameter $l/H_p$. Consider for example the discussion in \cite{bono_02}. Increasingly, Cepheids and related variables are used to investigate stellar populations, see, for example, \cite{marconi12}.  Hence, a proper assessment of the validity of conclusions is becoming both more important and, because of the possible influence of differing chemical abundances on the mixing length and other parameters, more difficult.
 
Another emerging area of research is the detection of multiple periodicities present in some of these objects due to the advent of 
long, uninterrupted observations of high quality with the CoRoT and Kepler missions. The detection of 
nonradial modes, e.g. \citet{gugg12}, amounts to new tasks for nonlinear modelling and in particular leads to the need 
for multidimensional simulations.

Let us also point out the usefulness of multidimensional Cepheid and RR Lyr studies for investigation of the atmospheres 
of these objects and the implications of such work. Whereas in solar and in general stellar physics use is increasingly 
made of multidimensional model atmospheres for abundance analysis and other types of research, e.g. \citet{aspl09}, 
static model atmospheres are predominantly used in work about Cepheids with only few exceptions, e.g. \citet{nard04}.
An appropriate treatment of the so-called projection factor which relates actual pulsation velocities to those 
observed turns out to be of importance in establishing the rotation law of the galaxy and issues regarding Cepheid 
distance scale calibration (cf.\ \citealt{nard09}). Quite generally, the increasingly sophisticated analysis of 
light curves of Cepheids and RR Lyr stars, e.g. \citet{kanb10}, should be backed up by a comparable sophistication 
of atmospheric models, therefore possibly hydrodynamical and, if convection is important, multidimensional models.

With these long-term goals in mind we present in this paper 2D hydrodynamical models of a Cepheid. As discussed already 
in paper I, the vastly different spatial scales and therefore stringent resolution requirements make this a
computationally quite intensive task even in 2D. In the following we investigate properties of the He~{\sc ii} partial ionisation zone, in 
particular issues regarding the energetics of the convection and the overshoot zone and implications for modelling of convection done in the traditional way 
in this area (models of such a sort that they fit into the framework of the geometrically
one dimensional hydrodynamic modelling of radially pulsating stars), and we discuss issues of the atmospheric structure 
and numerical complications which are not so severe in low resolution and show their full relevance only in high 
resolution simulations.

The remainder of this paper is organised as follows. In Sect.~\ref{sec:models} we describe the physical 
parameters of the Cepheid which we model with our simulations. 
The analysis of the enthalpy flux in and around the partial ionisation zone of He~{\sc ii} and the
mechanical work performed by the pulsation over a complete cycle as well as the comparison of these 
quantities with predictions from time-dependent convection models is presented in Sect.~\ref{sec:pulsation}.
In Sect.~\ref{sec:discussion} we provide a discussion of our results and present our conclusions.

\section{Models}   \label{sec:models}
 
The model we discuss in this paper corresponds to a fundamental tone Cepheid. Its parameters have already been given in paper I. We repeat them here for convenience.
 
Our star has an effective temperature  of $T_{\mathrm{eff}}=5125~\mathrm{K}$ and a gravitational acceleration at its surface of $\log g \sim 1.97$. Its luminosity is $L  \sim 913~L_{\odot}$, the initial radius is
$R=26.8~\mathrm{Gm}$ (which is about $38.5~R_{\odot}$), and the total stellar mass amounts to $5~M_{\odot}$. 

The chemical composition is characterized by $X=0.7$ and $Z=0.01$, the mix for $Z$ being taken from \citet{grev93}. The outer 11.3 Gm are modelled. As a result the pulsation period $P$ measured from our simulations is about 3.85 days which is slightly shorter than if one included the entire star in the computational domain.

The present work is based on a model whose sector has an opening angle of  $10^{\mathrm{o}}$.
The vertical cell size  ranges from
$0.47~\mathrm{Mm}$ at the top to 124 Mm at the bottom.  Further details on the numerical parameters  for the model and on the numerical methods used to solve the dynamical equations as well as a description of the dynamical equations themselves are given in paper~I.

\section{Pulsation and convection}    \label{sec:pulsation}

In this section we discuss properties of the He~{\sc ii} of the model just mentioned. This work is based on data covering ten useful pulsation periods after model relaxation. Averaging over ten periods allows for a sufficient removal of special properties of individual periods. After all, a problem typically experienced when running 2D convection simulations  the  small number of plumes present at any given time unless the domain is very broad. This makes for poor statistics compared to 3D runs with similar horizontal extent. The He~{\sc ii} zone is well resolved in each of the models we consider in the following.

\subsection{Comparison to traditional convection models}   \label{sec:compare model}

\subsubsection{Definition of the fluxes}   \label{sec:flux def}

Given the importance of convection modelling in traditional (1D) Cepheid calculations naturally great efforts have been undertaken to derive such models. The most frequently used models are those of \citet{kuhf86} and of \citet{stell82} as well as a number of variants based on them. 

Convection modelling in the sense of Kuhfu{\ss} has been adopted with modifications in the code described by \citet{wufeu98}.
Variants of the Stellingwerf model and the Kuhfu{\ss} model have been used in codes such as those of \citet{bono_st94} and the
Florida-Budapest cooperation (e.g \citet{buch_koll97}) and also in the more recent code described in \citet{smolec_mo08}.
 
These convection models introduce one additional time-dependent differential equation for the turbulent kinetic energy density, $e_{\rm t}$, plus algebraic recipes for various physical quantities. More elaborate convection models using the Reynolds stress approach have been developed, in particular those of \citet{canuto97} and \citet{canuto_dubo98}, but they have not yet been integrated into $1\mbox{D}$ nonlinear stellar pulsation codes. The comparably complex model of \citet{xiong97} has been used to study the pulsational stability of red giants, but only at the level of providing the background model for linear, nonadiabatic stability analyses (see \citet{xiong98} and \citet{xiong07}). We therefore concentrate on examining the prescriptions of the Kuhfu{\ss} and Stellingwerf models.
 
One of the differences between these models is the way in which the convective flux is expressed from the basic quantities, i.e., those derived from the Navier-Stokes equations plus $e_{\rm t}$.
Errors in the prediction of the convective flux may stem from two sources: from errors in 
$e_{\rm t}$ itself (and therefore the basic equation for this quantity) and from an inadequate algebraic expression used to calculate the convective flux $F_{\rm c}$. In the tests performed here we compute $e_{\rm t}$ from our $2\mbox{D}$ simulations and thus we test the algebraic form for $F_{\rm c}$. For two reasons this is the proper way to proceed in the present context. The first is a physical one and the other is of a primarily numerical nature.

The \textit{physical} reason is conected with the formulation of the time dependent convection (TDC) equation.
In the form given by \citet{buch_koll_00} it reads
\begin{equation}
\label{eq:eturb}
\partial_t{e_{\rm t}} + (p_{\rm t}+p_{\nu})\partial_tV =
  -\frac{1}{\rho r^2}\partial_r(r^2 F_{\rm t}) + C\mbox{,}
\end{equation}
where $p_{\rm t}$ is the turbulent pressure and $p_{\nu}$ is the viscous eddy pressure. 
$F_{\rm t}$ is the flux of turbulent kinetic energy and $C$ is a coupling term of the form 
\begin{equation}
C = S - \epsilon,
\end{equation}
where $S$ denotes a source term and $\epsilon$ is the dissipation rate of the (turbulent) kinetic energy. Both the viscous eddy pressure $ p_{\nu} $ and the dissipation rate $ \epsilon $ concern actions on a very small scale and in actual practice cannot be
computed without model assumptions concerning just those small scale processes. From the standpoint of $2\mbox{D}$ or $3\mbox{D}$
modelling that means, however, that these quantities cannot readily or unambiguously be recovered from the numerical simulations.
After all, the latter contain information about the resolved scales only. The hope and expectation (which to a considerable extent
was supported  e.g. by simulations of solar granulation) in connection with such models is that they deliver useful and reliable
answers even if the parameters (Reynolds number, Rayleigh number, etc.) in the numerical models differ widely from those found in
actual objects (cf.\ \citet{kup09}). At the same time it should be kept in mind that Eq.~(\ref{eq:eturb}) in itself is
phenomenological only. It is in at least one way connected with simplified expressions for $\epsilon$ and connected to the Kolmogorov
dissipation picture as described in \citet{buch_koll_00}. However, Eq.~(\ref{eq:eturb}) is not built into the multidimensional simulations.

Considered from a more basic point of view, Eq.~(\ref{eq:eturb}) is not conservative. Moreover it cannot be brought into that form with the approximate associated
expressions for $\epsilon$ and $S$ . The physical reason for this is that turbulent kinetic energy
considered on its own is simply not a conserved quantity: there has to be a source  of turbulent kinetic energy to prevent it from
decaying. In the models used in practice the source and dissipation terms are, from a quite fundamental point of view, not provided in conservation form. On the other hand, our numerical methodology in ANTARES rests on conservation form. This is for the very good reason that
in hydrodynamics only conservation form can be expected to deliver correct answers for shock speeds and the like (cf.\ \citet{trang09}). Nevertheless, even if we chose to solve Eq.~(\ref{eq:eturb}) alongside, it would not have the same, or any, backreaction on our simulations because it is not one of the dynamic equations we solve. It would therefore concern a quantity different from that one appearing in the TDC models despite of a formal, superficial analogy.

Therefore, similar to \citet{gast_L11}, we undertake a direct comparison of the convective flux between the numerical hydrodynamical simulations and its approximation provided by the TDC models. Each of the TDC based fluxes has the form
\begin{equation} \label{eq:flux}
	F_{\rm c,\ast} = \alpha_{\rm c,\ast}\eta_{\ast}\phi_{\ast}.
\end{equation}
Here, $\ast$ is $\mbox{\rm KF}$ for Kuhfu{\ss} and $\mbox{\rm SW}$ for Stellingwerf. $\eta_{\ast}$ is a correction factor whose purpose is to account for radiative energy losses of the convective elements. For the Kuhfu{\ss} and the Stellingwerf case a slightly different functional form is used (see below). $\eta_{\ast}$ contains, among others, the Peclet number ${\rm Pe_{\ast}}$, which, again, is calculated somewhat differently in the Kuhfu{\ss} and in the Stellingwerf setting.

Our use of $\alpha_c$ complies with \cite{gast_L11}. Note that e.g.  
\cite{buch_koll_00} use a somewhat different way to write the factor.

The functions $\phi_{\ast}$, $\alpha_{\rm c,\ast}$ and $\eta_{\ast}$ depend on the depth. 
In the \textit{Stellingwerf} model $\phi_{\rm SW}$ is computed from
\begin{equation} \label{eq:phist}
	 \phi_{\rm SW}=  
	   \Theta \cdot \rho \cdot e_{\rm t} \cdot \mbox{sign}(\nabla_{\rm supad})
	  \cdot \vert \nabla_{\rm supad} \vert ,
\end{equation}
where  $\Theta = \sqrt{c_{\rm p} \cdot T / \nabla_{\rm ad}} $. It furthermore adopts
\begin{equation} \label{eq:eta_SW}
	\mathcal{\eta}_{\rm SW} = \frac{1}{1+\alpha_{\rm r} /{\rm Pe}_{\rm SW}},
\end{equation}
with a Peclet number computed from the following expression:
\begin{equation}
{\rm Pe}_{\rm SW} = H_{\rm p} \cdot \sqrt{e_{\rm t}}\cdot \frac{c_{\rm p}\cdot \kappa_{\rm ross} \cdot \rho^{2}}{\frac{16\sigma}{3}\cdot T^{3}}\mbox{.}
\end{equation}
This form was proposed in \citet{buch_koll_00} (see Sect.~5.3 therein) originally with the model of Kuhfu{\ss} in mind, although at this level the expressions for the convective flux have the same general structure (\ref{eq:flux}) for both the Stellingwerf and the Kuhfu{\ss} models. 
In (\ref{eq:eta_SW}), $\alpha_{\rm r} $ is a parameter, unknown from the outset, $\kappa_{\rm ross}$ 
is the Rosseland mean opacity, $\sigma$ the Stefan--Boltzmann constant and $T$ the temperature. 
Moreover, $c_{\rm p}$ is the usual heat capacity at constant pressure, $\rho$ denotes the density, $e_{\rm t}$ the turbulent kinetic energy density, and $H_{\rm p}$ stands for the pressure scale height. Finally, $\nabla_{\rm supad}$ is the superadiabatic temperature gradient, 
\begin{equation}
\nabla_{\rm supad}=\nabla-\nabla_{\rm ad},
\end{equation}
and $\nabla$ denotes the actual temperature gradient ${\rm d\, ln}\, T / {\rm d\, ln}\, P$ while 
$\nabla_{\rm ad}$ is the adiabatic temperature gradient.

For the \textit{Kuhfu{\ss}} model we have
\begin{equation} \label{eq:phiku}
	 \phi_{\rm KF} =  
	  c_{\rm p} \cdot \rho \cdot \sqrt{e_{\rm t}} \cdot \nabla_{\rm supad}.
\end{equation}
To demonstrate the effect of radiative losses on the Kuhfu{\ss} model if approximated as in
\citet{buch_koll_00} we introduce here a factor $\mathcal{\eta}_{\rm KF}$ which contains some parameter 
$\alpha_{\rm r}$. Again, it influences the efficiency with which the model diffuses surplus energy of convective elements and it is assumed to be of the form
\begin{equation} \label{eq:eta_KF}
	\mathcal{\eta}_{\rm KF} = \frac{\alpha_{\rm r} \cdot {\rm Pe}_{\rm KF}}{1+\alpha_{\rm r} \cdot {\rm Pe}_{\rm KF}}.
\end{equation}
The  Peclet number for the Kuhfu{\ss} model is computed from
\begin{equation}
\label{eq:Pe_KF}
  {\rm Pe}_{\rm KF} = \frac{\sqrt{0.4}\cdot\pi}{5}\cdot H_{\rm p} \cdot \sqrt{e_{\rm t}} \cdot
     \frac{c_{\rm p}\cdot{\rho}}{K_{\rm rad}},
\end{equation}
where $K_{\rm rad} = (16 \sigma T^{3}) / (3 \kappa_{\rm ross} \rho)$ is the radiative conductivity. This is the expression for ${\rm Pe}$ derived by \citet{canuto_dubo98} for the case that the dissipation rate $\epsilon$ 
is computed by means of a mixing length $l$ in the special case $l=H_{\rm p}$. Note that Eq.~(\ref{eq:eta_KF}) can be transformed into Eq.~(\ref{eq:eta_SW}): the role of $\alpha_{\rm r}$ is reversed between the two,
i.e.\
\begin{equation}
\alpha_{\rm r, KF} = \alpha_{\rm r, SW}^{-1}\, 5/(\sqrt{0.4}\,\pi)
\end{equation}
while
${\rm Pe}_{\rm KF} = (\sqrt{0.4}\,\pi/5)\, {\rm Pe}_{\rm SW}$. Consequently, all the differences between
Eq.~(\ref{eq:eta_SW}) and Eq.~(\ref{eq:eta_KF}) can be absorbed into the numerical value for $\alpha_{\rm r}$ 
and the two forms of computing $\eta$ are equivalent. Eq.~(\ref{eq:eta_KF}) has the advantage of a more
straightforward comparison with the model of \citet{canuto_dubo98} and extensions thereof in future work. 
We note that more detailed prescriptions to compute radiative losses within the Kuhfu{\ss} model itself have
been suggested by \citet{wufeu98}, which, however, quite closely couple the computation of the convective 
flux with the dynamical equation for $e_{\rm t}$. In this study we limit ourselves to the prescription
given by Eq.~(\ref{eq:flux}) which can readily be applied to both TDC models, as intended by \citet{buch_koll_00}.

Last but not least we also compute the convective flux from our simulations through
\begin{equation}   \label{eq:flux_sim}
F_{\rm c}=u_r'\cdot\left(\rho\cdot h\right).
\end{equation}
Here $u_r'=u_r-\overline{u_r}$ is the vertical convective velocity (the radial velocity 
after subtraction of the radial pulsational velocity) and $h$  the enthalpy.

\subsubsection{Flux properties and comparison}
\label{sec:flux_properties}

In order to compare the fluxes used in traditional Cepheid modelling with those from  our simulations and relating them to the actual convective fluxes as calculated form the simulations we have averaged the latter ones over ten useful pulsation periods. In this way we improve the statistics. After all, it is known that 2D simulations have disadvantages regarding statistics compared to 3D simulations due to the small number of convective plumes present at any given instance of time unless the domain is very wide. For each of the results we discuss in this section the fluxes have been averaged horizontally.

\begin{figure*}
\subfigure[\sffamily $F_{\rm c3}=u_r'\cdot\left(\rho h- \overline{\rho h}\right)$]
   {\includegraphics[width=1\linewidth]{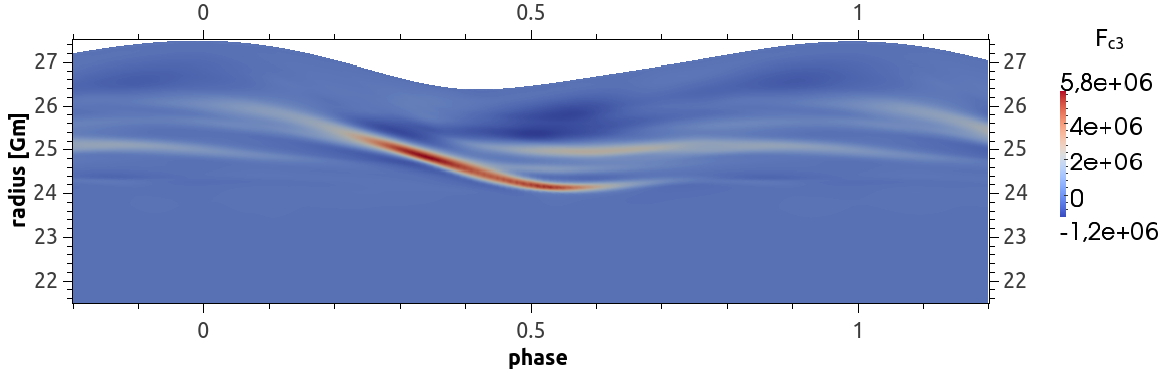}
   \label{pic:Fc3}}
\subfigure[\sffamily $F_{\rm c}=u_r'\cdot\left(\rho h\right)$]
   {\includegraphics[width=1\linewidth]{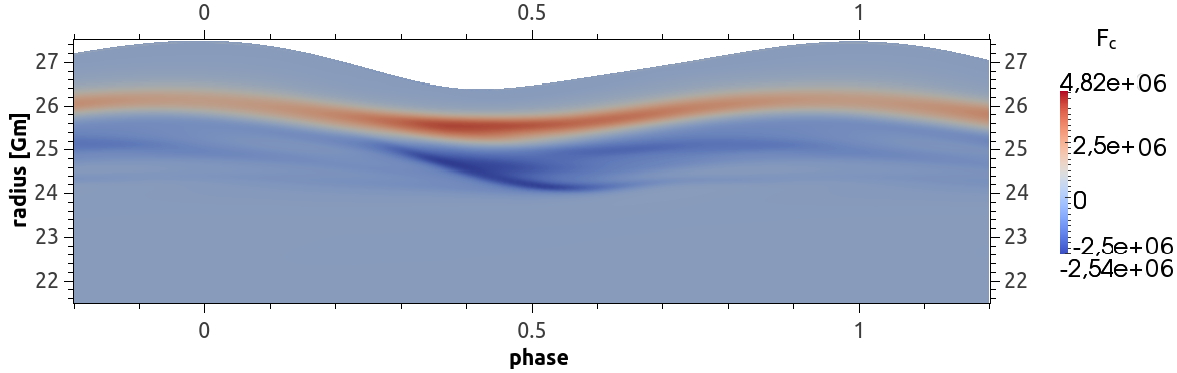}
   \label{pic:Fc2}}
\caption{The convective flux obtained from the simulation (in colour) after averaging over ten periods is plotted
here against phase. The vertical axis shows the radial extent of the model, while phase is plotted horizontally. In the upper picture the convective flux $F_{\rm c3}$ is computed only from the enthalpy perturbation. The units are $\rm J s^{-1} m^{-2}$ for both pictures.
Note the asymmetry of the radius as a function of phase, a feature well known through observations of Cepheids. Convective entities form high in the He~{\sc ii} partial ionisation zone are found to migrate downwards into the overshooting region.}
\label{fig:convFlux}
\end{figure*}

Fig.~\ref{fig:convFlux} displays the convective flux plotted against phase (averaged over ten periods). The variation of convection in intensity as a function of the phase is clearly visible. As Fig.~\ref{pic:Fc3} shows, convective activity (as witnessed by convective flux) commences its activity high up in the convectively unstable layer. Subsequently, plumes descend downwards. While their effect is still recognizable deep down (in the overshoot zone, in fact), new convective activity again sets in at the top of the He~{\sc ii} partial ionisation zone.

The two subfigures in Fig.~\ref{fig:convFlux} display two variants of the convective flux, namely $F_{\rm c3}=u_r'\cdot\left(\rho h- \overline{\rho h}\right)$, based on entropy perturbation against the mean, in Fig.~\ref{pic:Fc3}, and $F_{\rm c}=u_r'\cdot\left(\rho h\right)$ in Fig.~\ref{pic:Fc2}, where the total enthalpy enters (cf. \ref{eq:flux}). In the upper picture there is a noticeable orange-coloured streak, corresponding to  plumes directed downwards, in particular in the overshoot region. They have a large negative value of velocity, and $\rho h < \overline{\rho h}$ because they are hotter than the ambience in the overshoot region. This leads to a large positive value of $F_{\rm c3}=u_r'\cdot\left(\rho h- \overline{\rho h}\right)$, but to a negative value for $F_{\rm c}$.

\begin{figure*}
\begin{center}
\subfigure[\sffamily Comparision of $F_c$ and $\phi_{*}$]
   {\includegraphics[width=0.48 \linewidth,angle=0]{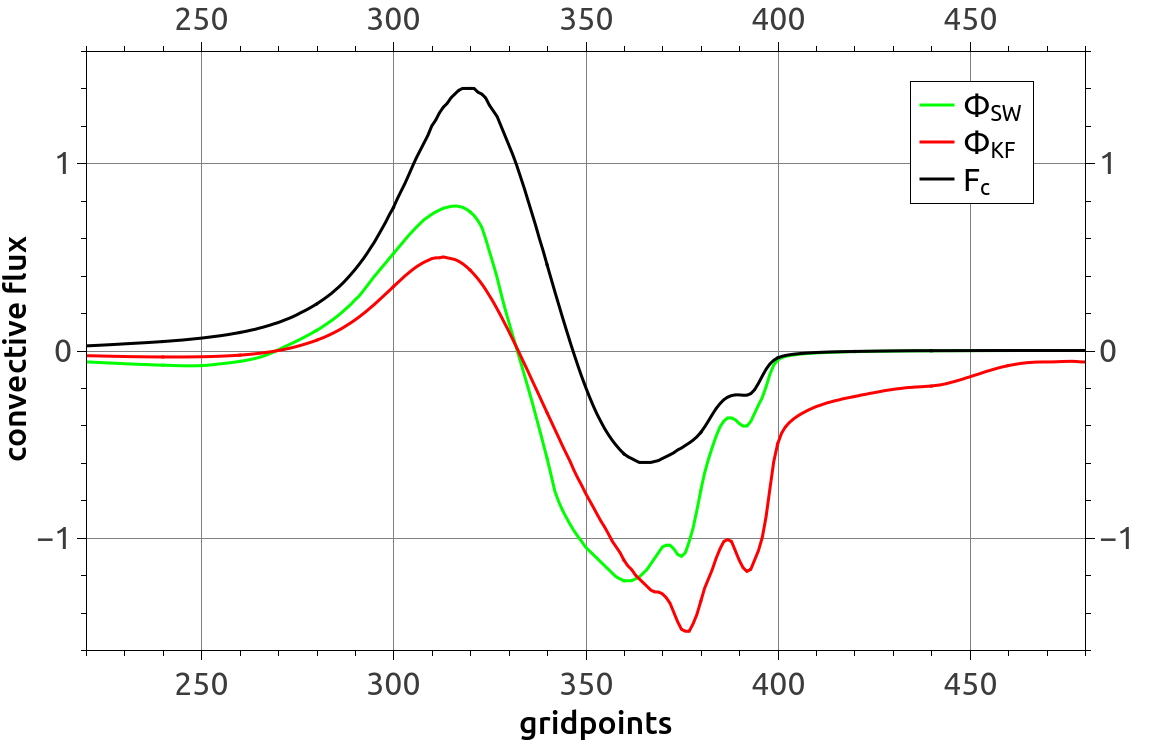}
   \label{pic:phiNoEta}}
   \hfill{}
\subfigure[\sffamily Comparision of $F_c$ and $\phi_\ast \cdot\eta_\ast$]
   {\includegraphics[width=0.48 \linewidth,angle=0]{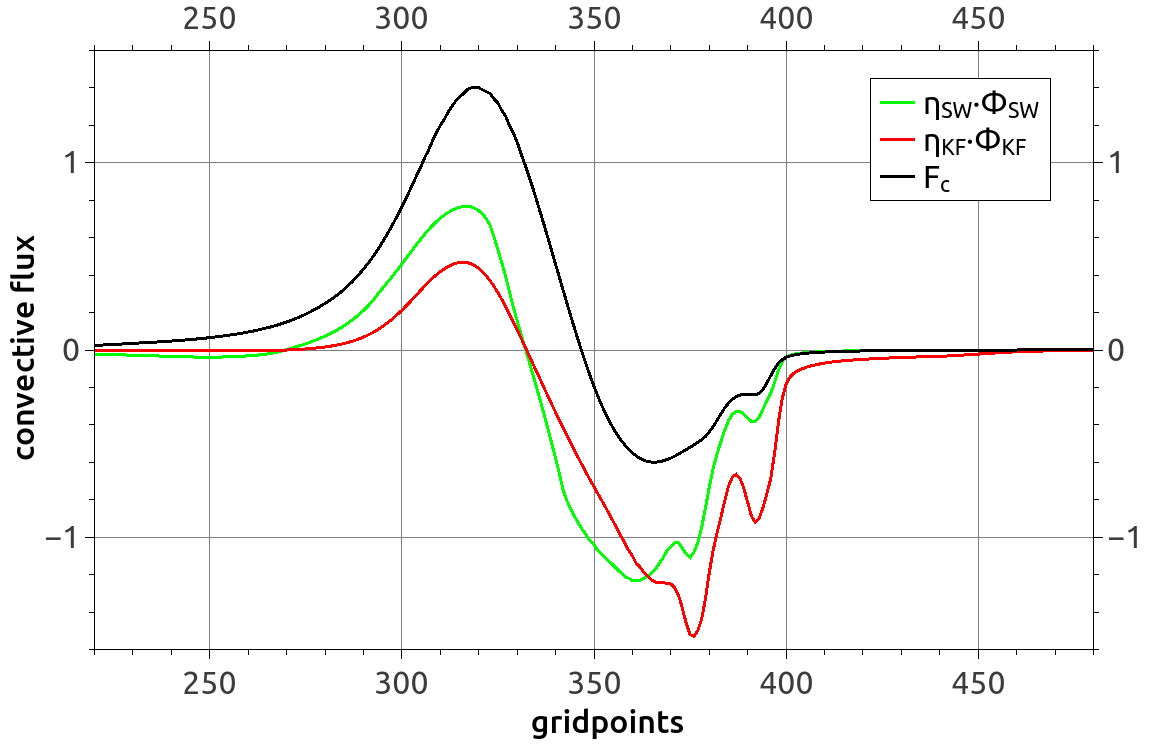}
   \label{pic:phiEta}} 
\caption{Mean convective flux in the simulation compared to the flux values for the Kuhfu{\ss} and Stellingwerf expressions for the convective flux. $\eta=1$ is assumed in the left panel whereas $\eta$ is determined according to Eq.~(\ref{eq:eta_SW}) and Eq.~(\ref{eq:eta_KF})    in the right panel. The top of the simulation box (at gridpoint 1, not in the range of the figure) is to the left of each panel.  
}
\label{fig:flux tot domain}
\end{center}
\end{figure*}

In Fig.~\ref{fig:flux tot domain} the mean convective flux $F_{\rm c}$ is compared to its approximation provided by the TDC models over the total computational domain. The left panel refers to the Kuhfu{\ss} and Stellingwerf fluxes without the correction factor, which amounts to $\eta_\ast=1$. In the panel to the right the correction factor is included. Similiar to the findings in \citet{gast_L11}, but less pronounced, the Kuhfu{\ss} formulation overestimates the convective flux in the overshooting region also in our case. Partial remedy can be achieved by including the Peclet-factor $\eta_\ast$ as is evident from Fig.~\ref{pic:phiEta}. The results based on the Stellingwerf model turn out to be only weakly dependent on the Peclet correction. -- The approximations also predict a slight negative overshooting above the convection zone. The transition from convection to overshooting occurs closer to the surface than for the convective flux proper. -- In all figures and tables hereafter the Peclet-factor will therefore be included.

\begin{figure*}
\begin{center}
\subfigure[\sffamily KF -- convection zone, $\alpha_r=0.3$]
   {\includegraphics[width=0.48 \linewidth,angle=0]{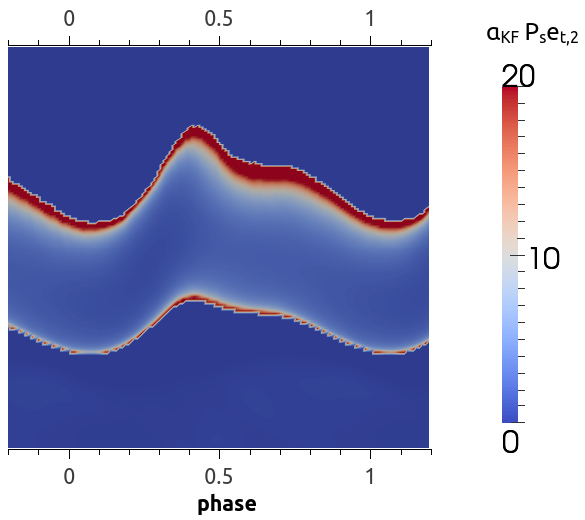}}
   \label{pic:KF-stripe}
   \hfill{}
\subfigure[\sffamily SW -- convection zone, $\alpha_r=0.1$]
   {\includegraphics[width=0.48 \linewidth,angle=0]{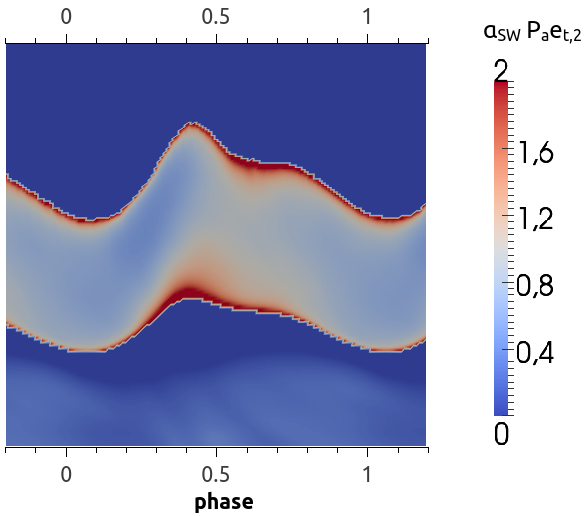}}
   \label{pic:SW-stripe} 
\caption{Horizontally averaged $\alpha_{\rm c}$-values for the Kuhfu{\ss} (left panel) and Stellingwerf (right panel) 
 expressions for the convective flux including $\eta_\ast $ and including the y-component of the turbulent kinetic energy $\left(e_{t,2}\right)$, indicated by a colour scale from blue (small values) to red (large values). The horizontal axis displays phase while depth is plotted along the vertical axis (range of radius from $24.5~\rm Gm$ to $26.4~\rm Gm$). An average over ten periods is shown.
 \label{fig:alpha_stripes}}
\end{center}
\end{figure*}

Fig.~\ref{fig:alpha_stripes} shows $\alpha_{c,\ast}$ as a function of phase for one full period (actually even from phase -0.2 to 1.2 for clearer  representation). $\alpha_{c,\ast}$ is obtained from $\alpha_{c,\ast}=F_{c,\ast}/\left(\phi_{\ast}\cdot\eta_{\ast}\right) $. The reddish vertical stripes represent the upper and lower limits, respectively, of the He~{\sc ii} convection zone proper quite well. A systematic variation of $\alpha_{\rm c,\ast}$ with both the depth in the convection zone and with pulsation phase is immediately apparent. Furthermore, comparing corresponding values from the convection zone and the overshoot zone, one can easily spot phases where the  $\alpha_{\rm c,\ast}$-values differ by a factor 5-15 and hence by an order of magnitude.

We therefore need to investigate the coefficients $\alpha_{c,\ast}$ as a function of phase separately averaged over the convection zone and the overshooting zone. To this end the following procedure has proved reliable and useful to determine whether a depth point belongs, at a given instant of time, either to the convection zone or the overshooting zone or to neither of them. 

A layer belongs to the \textit{convection zone} if, at this depth, more than $90$ percent of the horizontal points have a positive superadiabatic gradient, $\nabla-\nabla_{\rm ad}>0$. The gradients are computed from local values. 
Points below  the convection zone belong to the \textit{overshooting 
region} if they satisfy $\Gamma_3 < 1.51$  and fulfill, in addition, $\nabla-\nabla_{\rm ad}<-0.03$. This choice 
guarantees both for the convection zone and the overshooting zone that the upper and lower boundary of the 
convection zone are excluded. Near these boundaries the convective flux changes sign and hence is practically $0$ so that the ratio 
$F_{c,\ast}/\left(\phi_{\ast}\cdot\eta_{\ast}\right)$ requested for the evaluation of $\alpha_{\rm c}$ (cf. Fig.~\ref{eq:flux}) leads to meaningless results.
Moreover, in this region the enthalpy flux is small and the ratio $F_{c,\ast}/\left(\phi_{\ast}\cdot\eta_{\ast}\right)$ is physically unimportant. 
Inspection of the simulation data shows that the overshooting region defined in this way nicely contains 
the downwards penetrating plumes.

\begin{figure*}
\begin{center}
\subfigure[\sffamily KF -- convection zone]
    {\includegraphics[width=0.45\linewidth,angle=0]{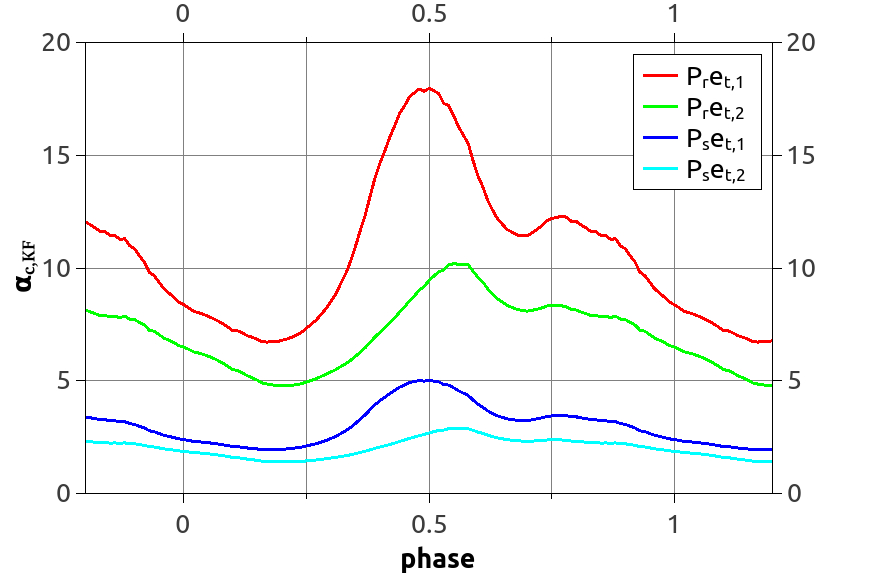}}
    \label{pic:KF-conv}
\hfill{}
\subfigure[\sffamily SW -- convection zone]
    {\includegraphics[width=0.45\linewidth,angle=0]{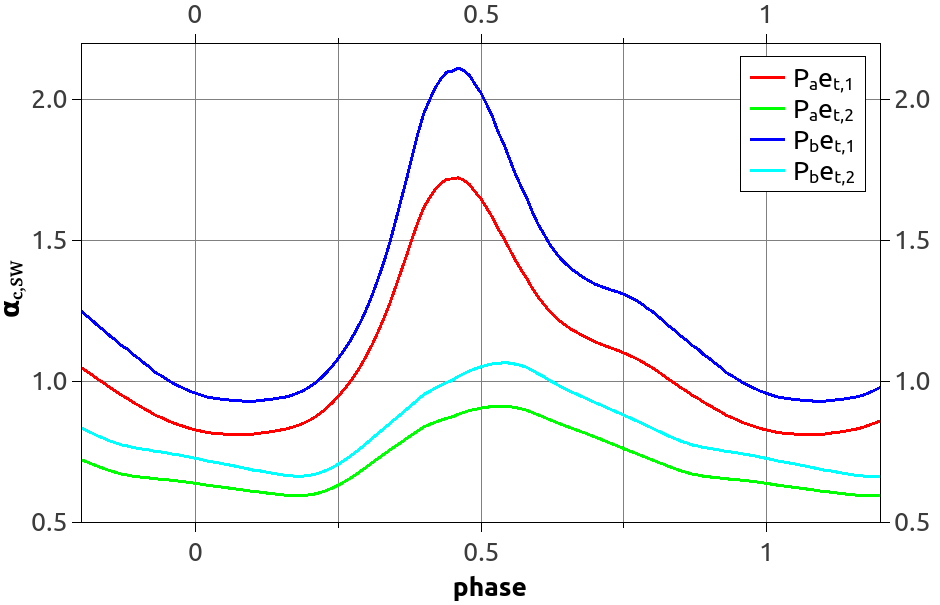}}
    \label{pic:SW-conv} 
\subfigure[\sffamily KF -- overshooting region]
    {\includegraphics[width=0.45\linewidth,angle=0]{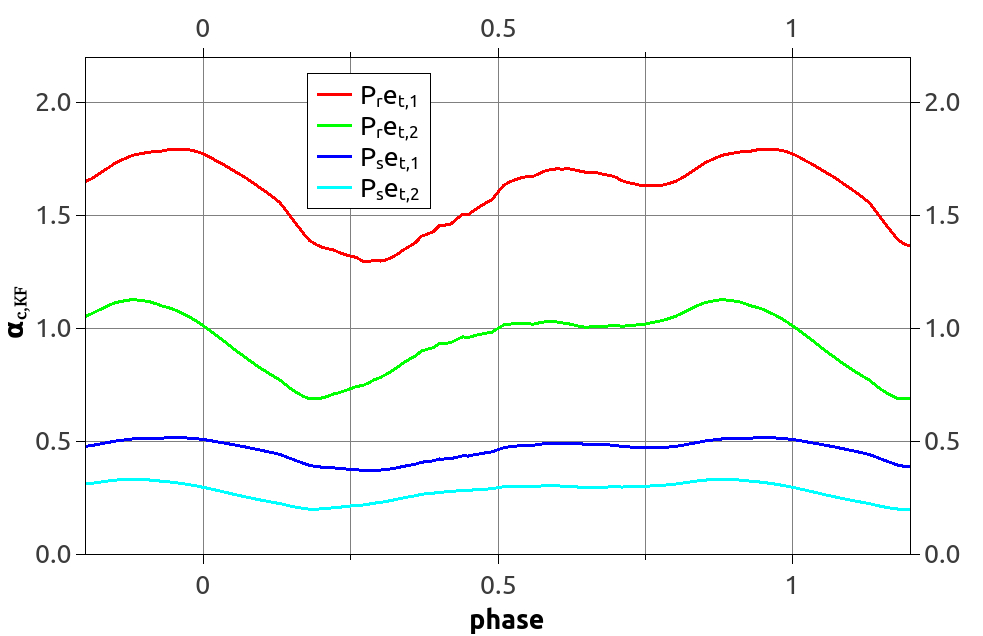}}
    \label{pic:KF-over} 
\hfill{}
\subfigure[\sffamily SW -- overshooting region]
    {\includegraphics[width=0.45\linewidth,angle=0]{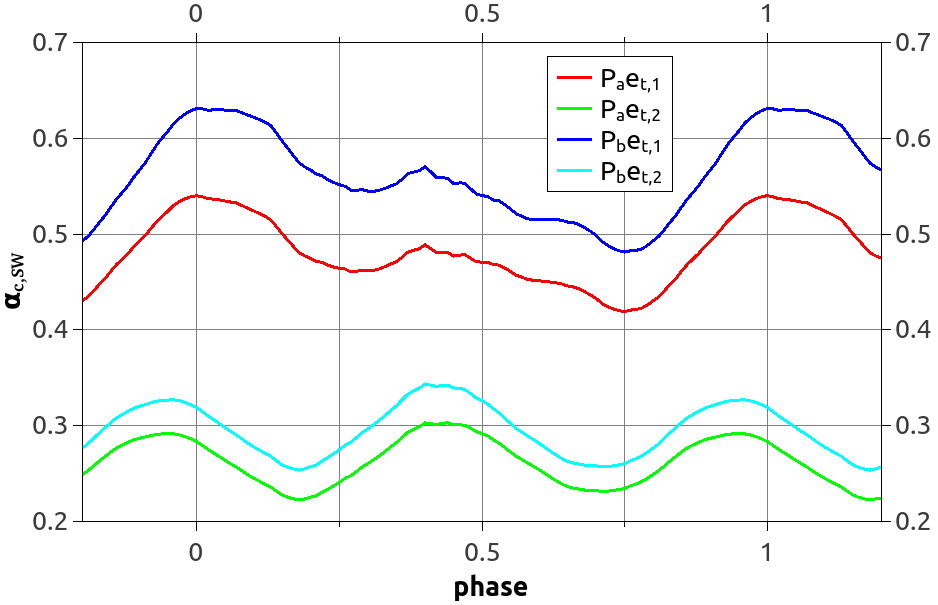}}
    \label{pic:SW-over} 
\caption{The parameter $\alpha_{\rm c}$ as a function of time: horizontal averages over
 $\sim$1 period. In the different panels KF denotes results from testing the Kuhfu{\ss} model
and SW results from testing the Stellingwerf model. Letters r, s, a, and b after P in the designation identify different values of $\alpha_{\rm r}$. The results denoted by $e_{\rm t,1}$ include only the vertical velocity in the computation of the turbulent energy while for results denoted by $e_{\rm t,2}$ the horizontal velocity was included as well. \label{fig:alpha01}}
\end{center}
\end{figure*}

\begin{figure*}
\begin{center}
\subfigure[\sffamily KF -- convection zone]
    {\includegraphics[width=0.45\linewidth,angle=0]{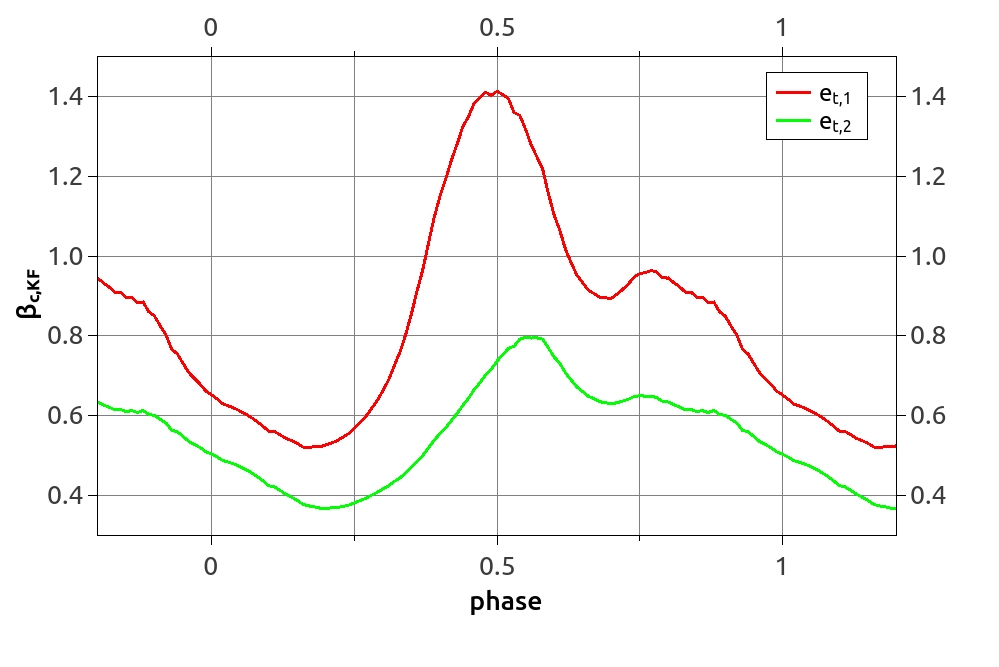}}
    \label{pic:be-KF-conv}
\hfill{}
\subfigure[\sffamily SW -- convection zone]
    {\includegraphics[width=0.45\linewidth,angle=0]{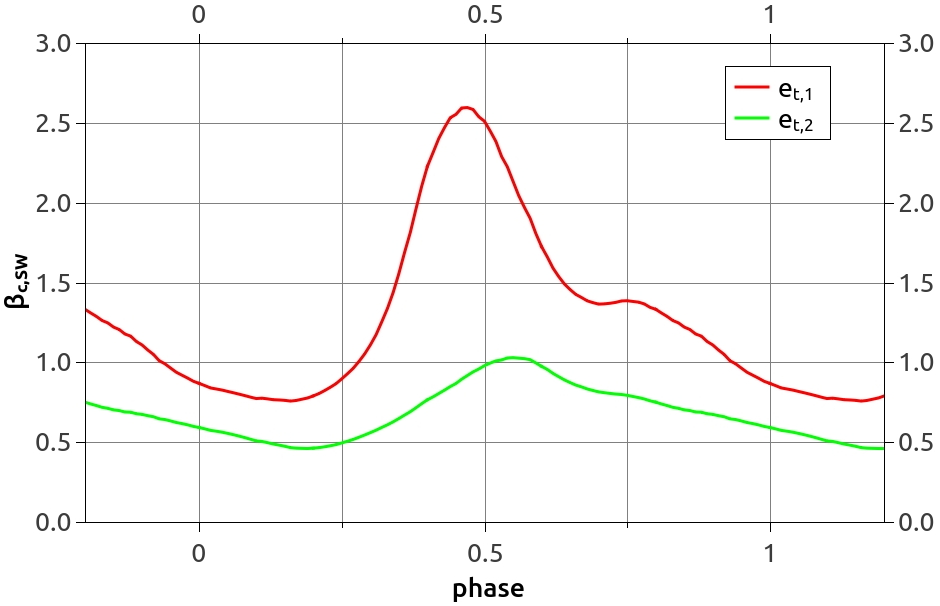}}
    \label{pic:be-SW-conv} 
\subfigure[\sffamily KF -- overshooting region]
    {\includegraphics[width=0.45\linewidth,angle=0]{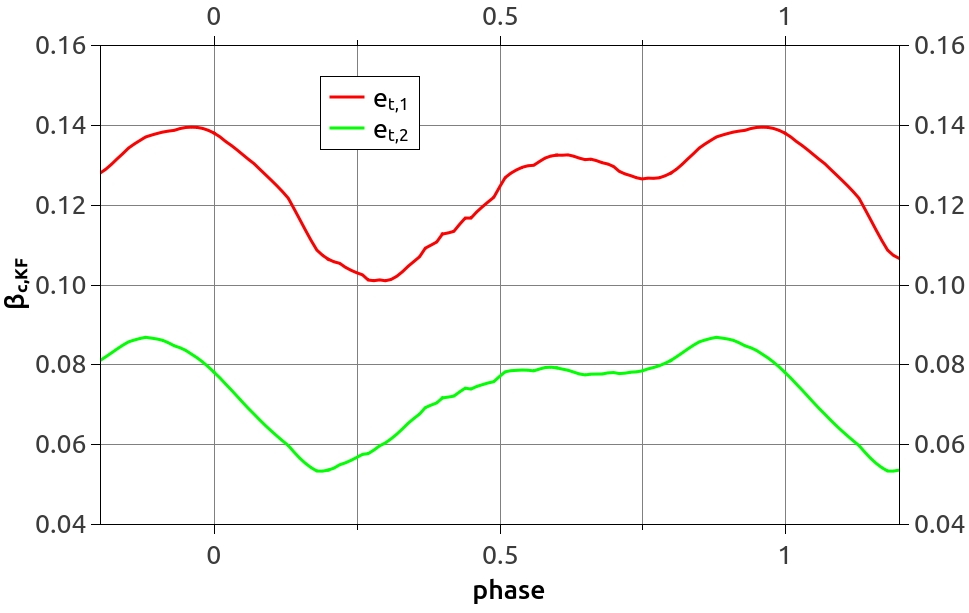}}
    \label{pic:be-KF-over}
\hfill{}
\subfigure[\sffamily SW -- overshooting region]
    {\includegraphics[width=0.45\linewidth,angle=0]{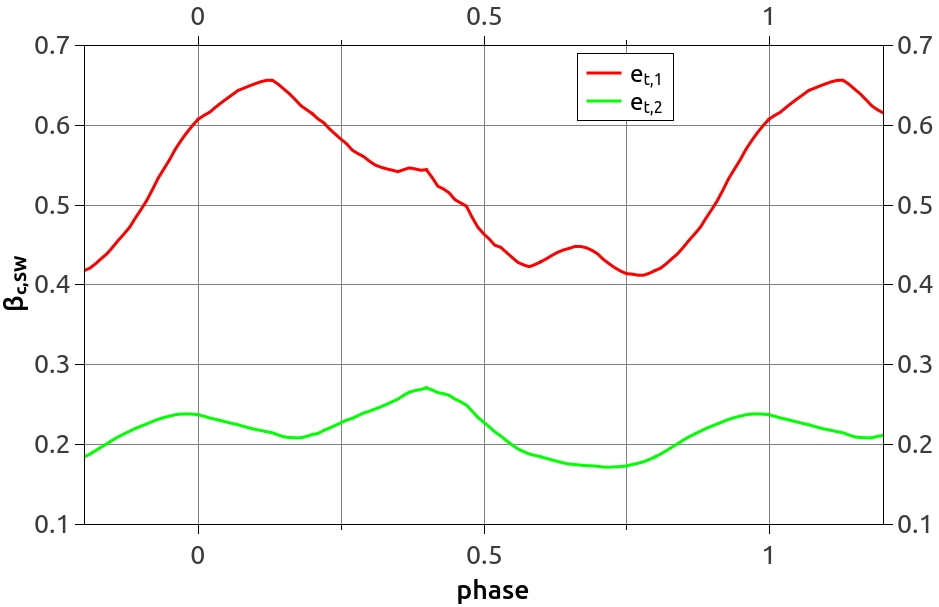}}
    \label{pic:be-SW-over} 
\caption{The parameter $\beta_{\rm c}$ as a function of time: horizontal averages for
 $\sim$1 period. In the different panels KF denotes results from testing the Kuhfu{\ss} model
and SW results from testing the Stellingwerf model. The results denoted by $e_{\rm t,1}$ include only the vertical velocity in the computation of the turbulent energy while for results denoted by $e_{\rm t,2}$ the horizontal velocity was included as well. \label{fig:beta01}}
\end{center}
\end{figure*}

\begin{table*}
\begin{minipage}{14cm}
\begin{center}
\begin{tabular}{ccrrrcrrc}
\hline
TDC & \multicolumn{2}{|c|}{variant} & \multicolumn{3}{|c|}{convection zone} &  \multicolumn{3}{|c|}{overshooting region}\\
\hline
    & designation & $\alpha_{\rm r} $        & $\alpha_{\rm c} $ & $\sigma$ & $\frac{\alpha_{{\rm max}}}{\alpha_{{\rm min}}}$ & $\alpha_{\rm c} $ & $\sigma$ & $\frac{\alpha_{{\rm max}}}{\alpha_{{\rm min}}}$  \\
\hline
\multirow{4}{*}{KF}
          & $\rm P_{\rm r}$ $e_{\rm t,1}$ & 0.08 & 11.2 & 31 & 2.8 &  1.59 &   10 & 1.4   \\ 
          & $\rm P_{\rm r}$ $e_{\rm t,2}$ & 0.08 &  7.2 & 22 & 2.3 &  0.95 &   14 & 1.7  \\
          & $\rm P_{\rm s}$ $e_{\rm t,1}$ & 0.30 &  3.2 & 30 & 2.7 &  0.46 &   10 & 1.4  \\ 
          & $\rm P_{\rm s}$ $e_{\rm t,2}$ & 0.30 &  2.1 & 22 & 2.2 & 0.28 & 14 & 1.7   \\ 
\hline
\multirow{4}{*}{SW} 
          & $\rm P_a$ $e_{\rm t,1}$ & 0.10 &  1.13 & 26 & 2.2 &  0.48 &   7 &  1.3  \\ 
          & $\rm P_a$ $e_{\rm t,2}$ & 0.10 &  0.74 & 15 & 1.5 &  0.26 &  10 &  1.4  \\ 
          & $\rm P_b$ $e_{\rm t,1}$ & 0.25 &  1.33 & 28 & 2.3 &  0.55 &   8 &  1.3  \\ 
          & $\rm P_b$ $e_{\rm t,2}$ & 0.25 &  0.84 & 16 & 1.6 &  0.30 &  10 &  1.4  \\
         
\hline
\end{tabular}
\end{center}
\caption{$\alpha$-values for various convection models: the designation indicates whether 
the $x$-component only is used for evaluating turbulent kinetic energy ($e_{\rm t,1}$) or whether 
also the $y$-component is used ($e_{\rm t,2}$). The letters r, s, a, and b in $\rm P_{\rm r}$, $\rm P_{\rm s}$, etc.\
refer to the value of $\alpha_{\rm r} $ adopted (listed in the next column). 
$\alpha_{\rm c} $ is the mean value, $\sigma$ the standard deviation in percent of the mean value 
(averaged each over ten periods). The ratio $\alpha_{{\rm max}} / \alpha_{{\rm min}}$ is a measure 
for the total amplitude of the variation of $\alpha_{\rm c} $ over one period.} 
\label{tab:alphas}
\end{minipage}
\end{table*}

\begin{table*}
\begin{minipage}{14cm}
\begin{center}
\begin{tabular}{ccccccccccccccccc}
\hline
TDC  & \multicolumn{3}{|c|}{convection zone} &  \multicolumn{3}{|c|}{overshooting region}\\
\hline
    &  $\beta_{\rm c} $ & $\sigma$ & $\frac{\beta_{{\rm max}}}{\beta_{{\rm min}}}$ & $\beta_{\rm c} $ & $\sigma$ & $\frac{\beta_{{\rm max}}}{\beta_{{\rm min}}}$  \\
\hline
\multirow{2}{*}{} 
 KF        & 0.56 & 23 & 2.2 &  0.07 &  13 & 1.6\\ 
 SW        & 0.72 & 25 & 2.3 &  0.22 &  14 & 1.7 \\ 
        
\hline
\end{tabular}
\end{center}
\caption{$\beta_{\rm c}$-values for various convection models  ($ \beta_{c,KF}\approx \alpha_{c,KF} \cdot \alpha_r$ and  $ \beta_{c,SW}\approx \alpha_{c,SW}/ \alpha_r$). Peclet factors contain both the  $x$- and $y$-direction contributions to the turbulent velocity. For further details see text. }  \label{tab:betas}
\end{minipage}
\end{table*}

\subsubsection{Domains of convection}
 
Given the fundamentally different nature of the convection zone proper and the overshooting region it is mandatory to investigate from the outset the achievements of the convection models for those two zones separately. To obtain a basis for those comparisons we have averaged data, which themselves are horizontal averages at each instant of time, over ten periods as already explained in Sect.~\ref{sec:flux_properties}.

The comparison of the convective flux obtained from the simulations with the TDC prescriptions calculated
as described in Sect.~\ref{sec:flux def} is summarised in Fig.~\ref{fig:alpha01} and in 
Table~\ref{tab:alphas}. The meaning of the abbreviations designating the variants of the convection
models are given in the caption of Table~\ref{tab:alphas} and $\alpha_{\rm c}$ is determined 
via Eq.~(\ref{eq:flux}). The large differences between the convection zone and the overshooting zone clearly stand
out as does the strong dependency of the values for $\alpha_{\rm c} $ on the adopted value for 
$\alpha_{\rm r} $. In addition, there is a strong variation of $\alpha_{\rm c} $ with phase. 
The dependence of $\alpha_{\rm c}$ on $\alpha_{\rm r} $ is of opposite sense in the Kuhfu{\ss} 
and the Stellingwerf model, respectively. This is an immediate consequence of the different 
definitions of $\alpha_{\rm r} $ in both cases (see Sect.~\ref{sec:flux def}). 

The quality, i.e.\ the constancy with phase, of a certain set of $\alpha$'s may be judged either from 
their standard deviation $\sigma$ or from the ratio $\alpha_{{\rm max}} / \alpha_{{\rm min}}$ (Table~\ref{tab:alphas}).
In all cases the two criteria lead to the same assessment of quality. In addition, there is the possibility to 
either calculate the turbulent kinetic energy from the vertical velocity alone or from both velocity components.
While the second option may appear to be the more natural one, the first option is of interest because  
some authors (for instance \citealt{gast_L11}, specifically Eq.~(2) in their paper) consider the contribution from the 
vertical velocity component only. Our results indicate opposite behaviour in the convection zone and the overshooting 
region, respectively: in the convection zone, inclusion of both components seems to lead to better results,
whereas in the overshooting region inclusion of the vertical velocity component only seems to be preferable.
That is true both for the Kuhfu{\ss} and the Stellingwerf formulation. On the other hand, the inclusion of the 
Peclet correction factor below the overshooting zone $\eta$ seems to have a rather marginal influence on the quality of the results due to the small influence of the specific value of $\alpha_r$ on the results, since a change in $\alpha_r$ can effectively be compensated by a change in $\alpha_c$. This is an immediate consequence of, for example, Eq.~\ref{eq:eta_KF}, the relevant values of $\alpha_r$ (see Table~\ref{tab:alphas}) and the fact that in our case therefore $\alpha_r\cdot \rm{Pe} \ll 1$.

\subsubsection{A first analysis}

At first it might seem surprising that the inclusion of radiative losses does not significantly improve the predictions of the convective flux by the TDC models. The best fit numerical values for $\alpha_{\rm c}$ indeed change as a result of including a factor $\eta \neq 1$, but there is hardly any improvement, if we compare the results for the convection zone with the overshooting region.
However, throughout most of the pulsation cycle the Peclet number ${\rm Pe}$ is of order unity or even less in the 
entire spatial domain analysed in our study. Indeed, the horizontal average barely ever reaches the value of $0.75$. Thus, we are dealing with convection in the low efficiency limit in both cases, i.e.\ for the convection zone proper and the overshooting region. For the He~{\sc ii} zone most of the convective driving stems from the lowering of $\nabla_{\rm ad}$ rather than from the rather mild local minimum in $K_{\rm rad}$. Consequently, the correction due to radiative losses is similar in both regions and the correction factor 
$\eta_\ast$ cannot compensate the large difference found in the convective flux predictions for the
convective and the overshooting region. Moreover, as it can be more easily seen from the form
of $\eta_{\rm KF}$ in Eq.~(\ref{eq:eta_KF}), it is the product $\alpha_{\rm c} \alpha_{\rm r}$ only that enters 
into the calculation of the convective flux in the limit of small Peclet number. Thus, a change
in $\alpha_{\rm r}$ just implies a scaling of $\alpha_{\rm c}$ throughout the simulation domain, 
since in the end we obtain $\alpha_{\rm c}$ from the ratio of Eq.~(\ref{eq:flux}) and Eq.~(\ref{eq:flux_sim}).
Hence, the problem of too large an overshooting found for the TDC models cannot be accounted for by just including a model of heat losses but must have an additional, more important contribution. -- We also note that the inclusion of heat losses does at least prevent even more excessive overshooting underneath the He~{\sc ii} partial ionisation  zone which would otherwise be borne out by the Kuhfuss model. This follows from a comparison of  Fig.~\ref{pic:phiNoEta} and Fig.~\ref{pic:phiEta}. 

Possibly the disparity between the $\alpha_c$-factors applicable to the convection or overshoot zone, respectively, we derive here would turn out to be even worse if similar work as the present one would be performed in 3D instead of 2D. In the 3D case overshooting is usually
less pronounced (cf.\ \citealt{muthsam95}) than in 2D simulations with otherwise identical setup. We therefore 
expect an even larger ratio of the $\alpha_{\rm c}$ between the convection zone and the overshooting zone. This suggests that 
work in 3D should be undertaken in the future.

We caution that the findings of the discussion above  cannot be simply extrapolated to the convection zone caused by the 
partial ionisation of H~{\sc i} and He~{\sc i}. There, both the variation of $K_{\rm rad}$ and the values of ${\rm Pe}$ are much larger.

The inclusion of the horizontal component of the turbulent kinetic energy into the calculation of $e_{\rm t}$ and, through
Eq.~(\ref{eq:flux}), into $F_{\rm c}$ is more important, as is clearly demonstrated by the results in Table~\ref{tab:alphas}.
The improvement within the convection zone for both TDC models is about 30\%. But no such improvement is found for the overshooting
zone: indeed, the model predictions become slightly worse. This can be understood by means of the anisotropy of the flows: the ratio of horizontal to vertical kinetic energy is different for both regions, since vertical flow dominates inside the
convection zone, while horizontal flow dominates outside. As the TDC models, implicitly or explicitly, assume some -- usually
constant -- degree of anisotropy, this result demonstrates that this quantity is variable and in particular significantly different 
inside and outside the convection zone proper. Clearly, this is one route along which the current TDC models could be improved.

It is interesting to approach the question of constancy of coefficients also from a different side for the He~{\sc ii} convection zone of our model. It is a weak convection zone, and the Peclet numbers are small both in the convection zone proper and in the overshooting region.  

However, even if the anisotropy of the distribution of kinetic energy and the amount of overshooting could be taken care of, the problem of the variability of the quality of the models with phase would likely remain, unless the convection models could incorporate the feedback between the mean structure and the turbulent kinetic energy as well as the convective flux in a more self-consistent manner. 

The fact that in the weak He~{\sc ii} convection zone the parameters $\alpha_c$ and $\alpha_r$ enter only through
\begin{equation}
 \label{eq:beta}
 \beta_{c,\ast}:=\frac{F_c}{\Phi_\ast \cdot Pe_\ast }
\end{equation}
as $ \beta_{c,KF}\approx \alpha_{c,KF} \cdot \alpha_r$ (Kuhfu{\ss}) or  $ \beta_{c,SW}\approx \alpha_{c,SW}/ \alpha_{c,SW}$ (Stellingwerf), respectively, raises the
question whether treating $\beta_{c,\ast}$ in the same way as has been done with $\alpha_{c,\ast}$ (cf. Table~\ref{tab:alphas},
Fig.~\ref{fig:alpha01}) might not result in a  better constancy of the flux predicted by the Kuhfu{\ss} or the Stellingwerf
prescription. Table~{\ref{tab:betas}} and Fig.~\ref{fig:beta01} teach, however, that this is not the case. Rather, parameters such
as ratio of the $\beta_{c,\ast}$-coefficients in the convection and the overshoot zone, respectively, or the variances tend to be
worse than is true for the corresponding values based on the $\alpha_{c,\ast}$-coefficients.

We turn attention to work of  \citet{gast_L11}. Although they have performed their simulations for a case
of idealised microphysics and plane parallel (as opposed to spherical) geometry and differ, moreover, in the numerical setup, they
have also found a temporal modulation of the convective flux (which can easily be read from our Fig.~\ref{fig:convFlux}) and,
moreover, they have found that the convective flux predicted by the Stellingwerf model is in somewhat better agreement with the
simulations than those obtained from the Kuhfu{\ss} model. The identified shortcomings of the most common TDC models used in the
study of convection-pulsation coupling are thus quite likely of a generic nature and not specific to Cepheids of a very narrow part
of the HR diagram.

Concluding this section we want to consider the following aspect. The $\alpha_{\rm c,max} / \alpha_{{\rm c,min}}$ ratio refers to quantities integrated over the convection or overshoot zone. 
In no way could the predictions of the TDC models be interpreted in the \textit{pointwise} sense (this has also not been suggested by the authors, as we hasten to remark). In this case the differences between the model predictions and the simulations would be even larger (by two orders of magnitude). From that point of view the averaging procedure still does a remarkable job in dealing with the point values which, taken \textit{per se}, are practically meaningless.

\subsection{Energetics of the convection zone}\label{subsec:pdV}

\subsubsection{Basic quantities}\label{subsubsec:pdV_basic}

The energy equation of hydrodynamics (e.g. \cite{castor04}) can easily be cast in the following form useful for our present purpose:
\begin{align}\label{eq:ener_lag1}
&D_t\left(e+e_{kin}\right) + \left(p+e_{kin}\right)\divergence{\boldsymbol{u}}=  \\  \nonumber
&-e\divergence{\boldsymbol{u}}-\boldsymbol{u}\cdot \gradient{p}+
\divergence{\left(\kappa\gradient{T}+\boldsymbol{u}\overline{\sigma}\right)}+\rho \boldsymbol{g}\cdot \boldsymbol{u}
\end{align}  
or equivalently
\begin{align}\label{eq:ener_lag2}
&D_t\left(\frac{e}{\rho}+\frac{e_{\rm kin}}{\rho}\right)+\left(p+e_{kin}\right) D_t\left(\frac{1}{\rho}\right)=  \\  \nonumber
&\frac{1}{\rho}
\left(e_{kin}\divergence \boldsymbol{u}-\boldsymbol{u}\cdot \gradient p+\divergence\left(\kappa\gradient T +\boldsymbol{u}\overline{\sigma}\right)\right)+
g\cdot \boldsymbol{u}\mbox{.}
\end{align}  

Here, $e$ denotes the internal and $e_{kin}$ the kinetic energy density. $p$ is the (thermodynamic) pressure, $\rho$ the density, $\overline{\sigma}$ the stress tensor, $\kappa$ the conductivity,  $T$ the temperature and $\boldsymbol{g}$ the vector of gravitational acceleration. $D_t$ is the Lagrangian time derivative. Note that $pD_t(\frac{1}{\rho})$ is close to the expression $pdV$ of thermodynamics (instead of a change of volume there is now a rate of change, and we work with an energy density instead of an energy enclosed in a specific volume, so that a term $-e\divergence \boldsymbol{u}$ appears on the right hand side to account for the rate of change of the volume) in Eq.~(\ref{eq:ener_lag1}).

\begin{figure*}
\subfigure[\sffamily]
   {\includegraphics[width=0.45\linewidth,angle=0]{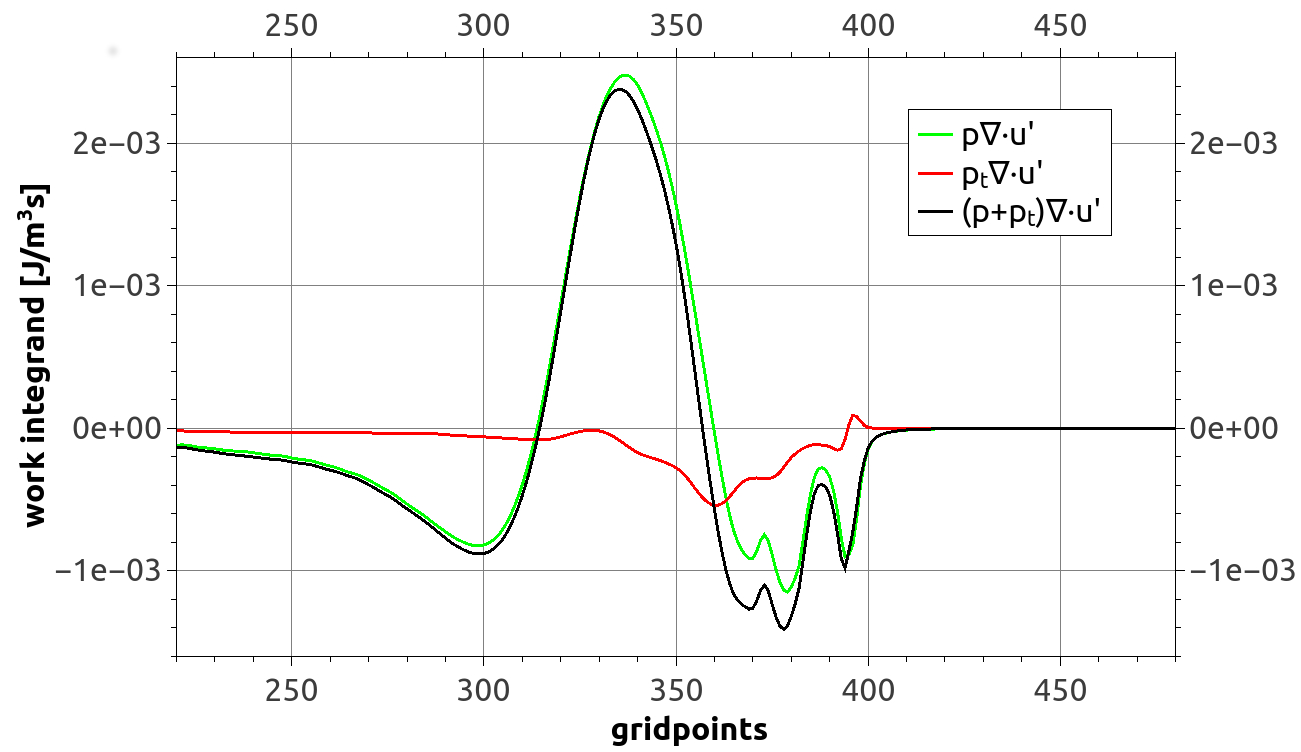}}
   \label{pic:W_pu_grid}
   \hfill{}
\subfigure[\sffamily]
   {\includegraphics[width=0.45\linewidth,angle=0]{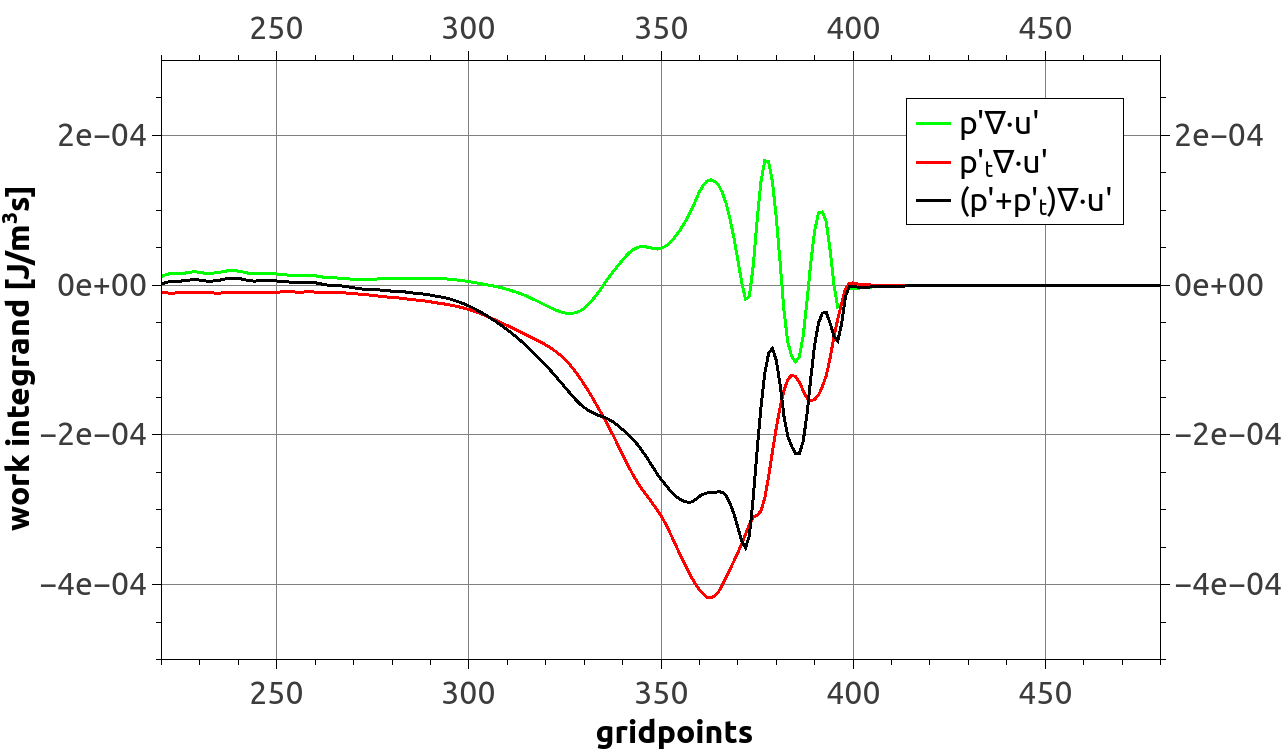}}
   \label{pic:Wm_pu_grid}
\caption{The work integrands for the fluctuating velocity field. In the left panel the pulsational component of each pressure term is included, while in the right panel only the pressure perturbations are used. The top of the simulation box is to the left of each panel.  \label{fig:WorkIntegrand}}
\end{figure*}

We now discuss several components of velocity and pressure, respectively, and their contribution to the so-called pdV-work.

To begin with, we denote by $\boldsymbol{\mu}$ the vector of momentum densities. In addition to the total velocity field $\boldsymbol{u}$, we consider the fluctuating velocity field
\begin{equation}
\boldsymbol{u}^{\prime}=\left(\frac{\mu_r}{\rho}-\frac{\bar{{\mu}}_r}{\bar{\rho}},\frac{\mu_\phi}{\rho}\right).
\end{equation}
The bar denotes horizontal averaging, so that  $\bar{{\mu}}_r$ stands for the horizontally averaged $r$-component of the momentum density. Since the relevant quantities ($\boldsymbol{\mu}$, $\boldsymbol{u}$, $p$, $\rho$) are combined in a nonlinear way   
it is not possible to define the horizontal means in a fully consistent manner unless one is ready to deal with a multitude of emerging mixed terms. The detailed way of 
averaging has, however, practically no influence on the results in the present case due to the smallness of the mixed terms.

Various components of the total pressure are of physical interest. $p$ denotes the thermodynamic pressure, as obtained from the equation of state. We then have
\begin{equation}
   p^\prime=p-\bar{p},
\end{equation}
the fluctuating thermodynamic pressure. As to the macroscopic pressure components, we consider
\begin{equation}
   p_t=\frac{\rho}{2}(u_r^2+u_\phi^2)
\end{equation}
and
\begin{equation}
   p_t^\prime=\frac{\rho}{2}((u_r^{\prime})^2+u_\phi^2).
\end{equation}
$p_t^\prime$ is often referred to as the turbulent pressure whereas $p_t$ contains also the radial pulsational component. The term \textit{turbulent}
pressure is not meant to imply any statement about the degree of turbulence of 
the simulated flow in terms of the Reynolds number or similar quantities. We rather use it in the way in which
it is used in the literature on one-dimensional simulations of stellar pulsation, denoting anything due to other than
the radially averaged motions. Modelling the turbulent pressure is a key issue in 1D recipes for the pulsation-convection interaction 
(e.g.  \cite{stell82}, \cite{yecko98}). 

To be more precise consider that some of these models, for example those in the papers just 
cited, also include an eddy viscous pressure as more closely described in \cite{gehmeyr92}. This eddy 
viscous pressure is assumed to be due to turbulence generated by the shear of the radial pulsation; cf. the discussion around Eq.~(4) and Eq.~(5)  in that paper. -- In multidimensional simulations there is neither a need nor a really convincing way to distinguish between those two kinds of turbulent pressure. Consequently, $p_t^\prime$ refers to the total turbulent pressure in such models.

 \subsubsection{The work components}\label{subsubsec:pdV_work_def}

The Lagrangian form of the continuity equation,
\begin{equation}
   D_t\rho = -\rho \divergence{\boldsymbol{u}} \mbox{,}
\end{equation}
combined with Eq. (\ref{eq:ener_lag2}) leads to work densities (per second) of the form pressure times divergence of velocity. The decompositions of pressure and velocity ($p=\overline{p}+p^{\prime}$, $\boldsymbol{u}=\overline{\boldsymbol{u}}+\boldsymbol{u}^{\prime}$) results in a number of work densities. Some of these are significant for the present discussion as inspection of data shows.

\begin{table*}
\begin{minipage}{16cm}
\begin{center}
\begin{tabular}{lrrrrrrrr}
\hline
region & \multicolumn{4}{|c|}{thermodynamic pressure} &  \multicolumn{4}{|c|}{turbulent pressure}\\
\hline
       & $\overline{W\left(p,\boldsymbol{u}\right)} $ & $\overline{W\left(p,\boldsymbol{u}^\prime\right)}$ &
         $\overline{W\left(p^\prime,\boldsymbol{u}\right)}$ & $ \overline{W\left(p^\prime,\boldsymbol{u}^\prime\right)}$ &
         $\overline{W\left(p_t,\boldsymbol{u}\right)}$ &  $\overline{W\left(p_t,\boldsymbol{u}^\prime\right)}$  &
         $\overline{W\left(p_t^\prime,\boldsymbol{u}\right)}$&$\overline{W\left(p_t^\prime,\boldsymbol{u}^\prime\right)}$\\
\hline 
convection zone     &  3.7e+27 & 2.6e+27 & 7.4e+25 & 4.0e+25 & -2.9e+27 & -7.5e+26 & -8.2e+26 & -5.1e+26\\ 
overshooting region &  9.3e+27 &-5.0e+26 & 5.8e+26 & 4.5e+26 & -2.5e+27 & -2.3e+27 & -2.9e+27 & -2.4e+27\\
        
\hline
\end{tabular}
\end{center}
\caption{Average values of the components of the work integral in various regions. Note that for the turbulent pressure all values are negative.}  \label{tab:pw-values}
\end{minipage}
\end{table*} 

The cycle averages over the horizontal average $\overline{p \cdot\divergence{\boldsymbol{u}}}$ of these work densities (\cite{buch_koll_00}) are the work integrands and express the amount of energy converted into pulsational energy. In Fig.~\ref{fig:WorkIntegrand} the work integrands are plotted for the fluctuating velocity. When comparing to Fig.~\ref{fig:flux tot domain} it can be seen that for the thermodynamic pressure dampening occurs at the top of the convection zone and that the transition from excitation to dampening is within the overshooting zone. The turbulent pressure has a dampening effect throughout where it is not small.  

These work densities can then be integrated, at any time, over the convection zone ($c$), the overshooting zone ($o$).
For the present purpose the definition of the convection zone is the same as given in  Sect.~\ref{sec:flux_properties}. The overshooting zone is now defined a little bit differently  from the one provided there. Namely, now  there is no gap between it and the convection zone: points below  the convection zone belong to the \textit{overshooting region} if they satisfy $\Gamma_3 < 1.51$  and fulfill, $\nabla-\nabla_{\rm ad}<0$ (instead of $\nabla-\nabla_{\rm ad}<-0.03$ in Sect.~\ref{sec:flux_properties}).
This results in terms such as, for example,  
$$W^o(p,\boldsymbol{u}^\prime)=\int_{o}p(x)\divergence{\boldsymbol{u}^\prime}(x)dx\mbox{,}$$ 
referring to the overshooting zone. 
In this way the rather different physical nature of the zones is taken into account. 

Within such an integral positive and negative contributions may cancel out very significantly. This can be seen by considering  the integrated \textit{modulus} of the pdV-terms such as
$$W_m^o(p,\boldsymbol{u}^\prime)=\int_{o}|p(x)\divergence{\boldsymbol{u}^\prime}(x)|dx$$
alongside with the original values. 

\begin{figure*}
\subfigure[\sffamily]
   {\includegraphics[width=0.45\linewidth,angle=0]{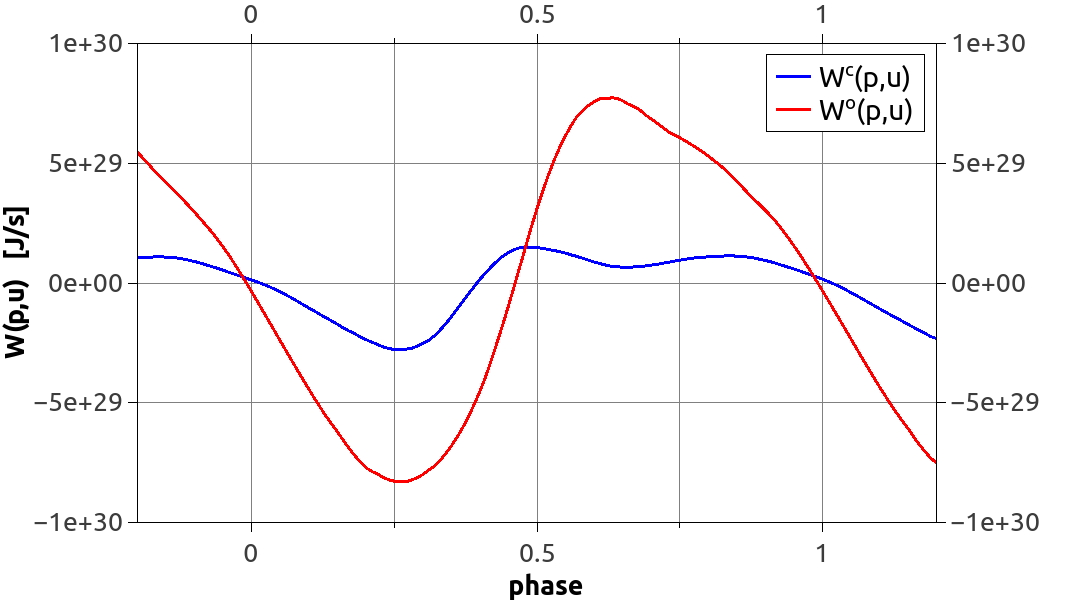}}
   \label{pic:Pu} 
   \hfill{}
\subfigure[\sffamily]
   {\includegraphics[width=0.45\linewidth,angle=0]{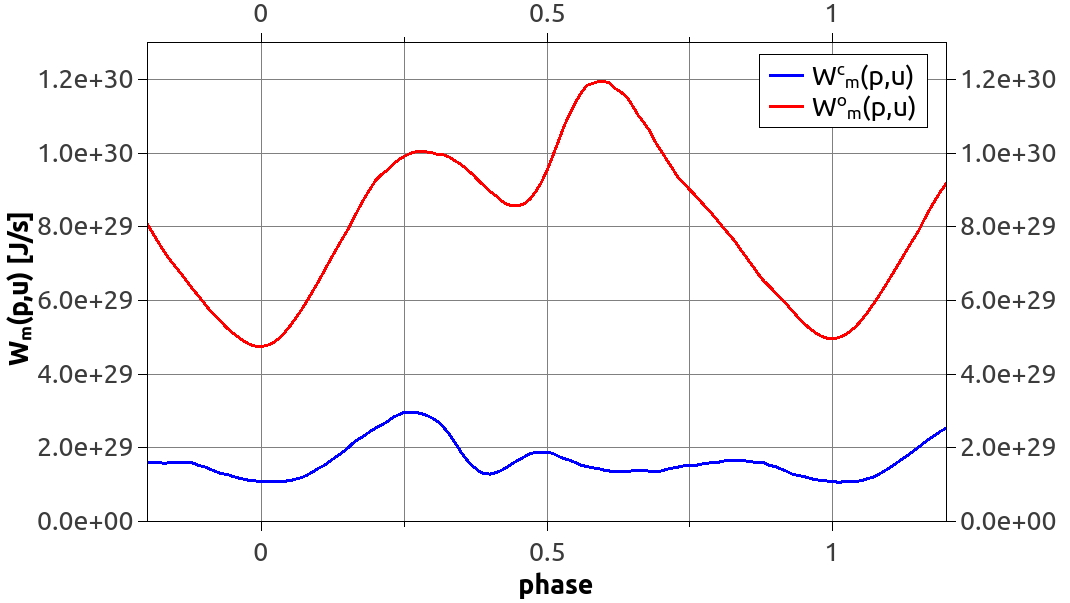}}
   \label{pic:mPu}
\subfigure[\sffamily]
   {\includegraphics[width=0.45\linewidth,angle=0]{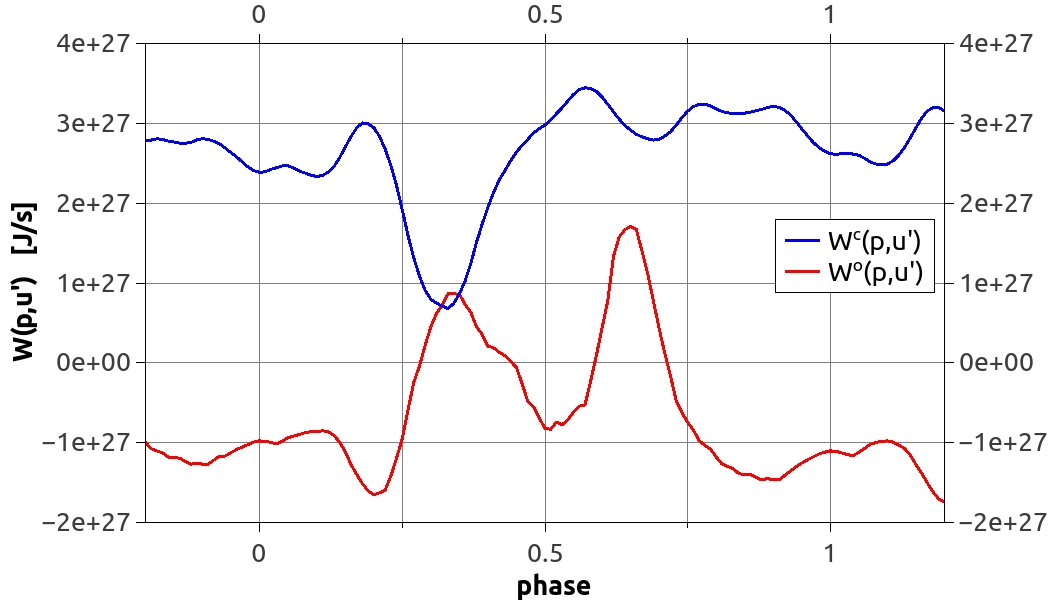}}
   \label{pic:P1u}
   \hfill{}
\subfigure[\sffamily]
   {\includegraphics[width=0.45\linewidth,angle=0]{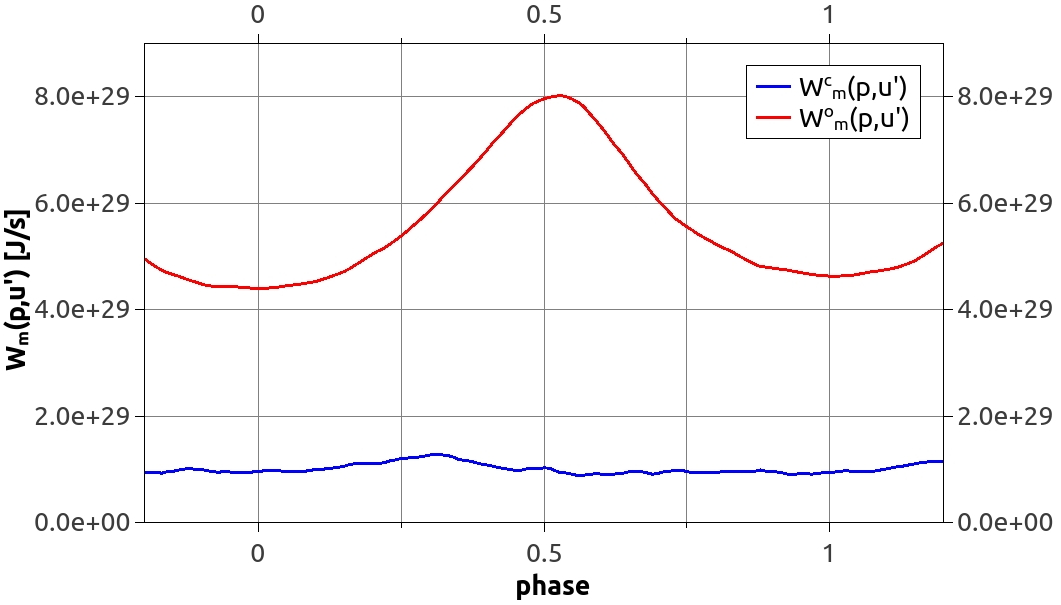}}
   \label{pic:mP1u}
\subfigure[\sffamily]
   {\includegraphics[width=0.45\linewidth,angle=0]{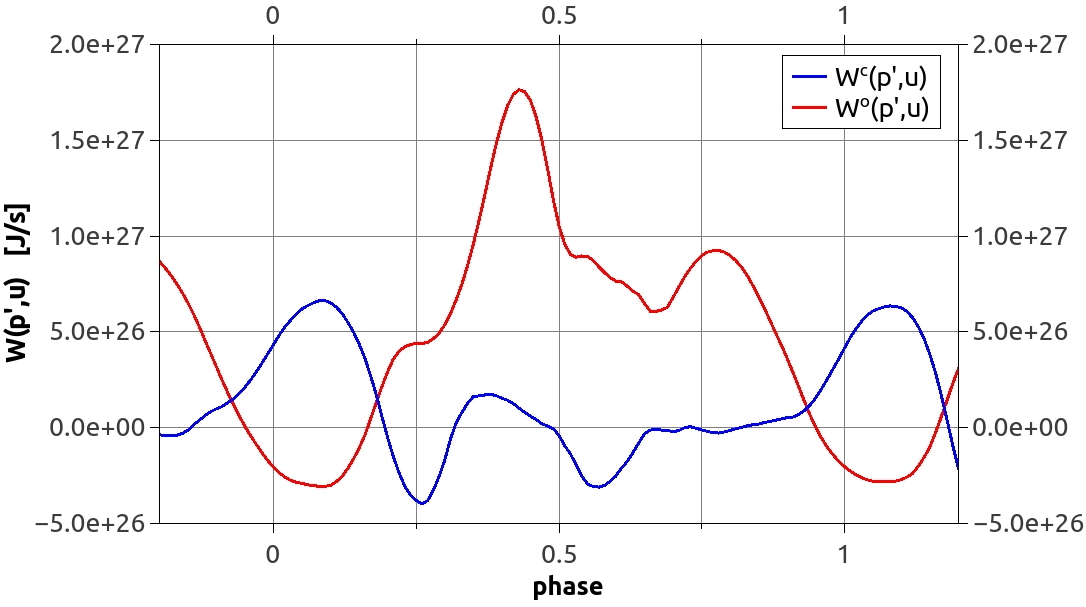}}
   \label{pic:Pu1}
   \hfill{}
\subfigure[\sffamily]
   {\includegraphics[width=0.45\linewidth,angle=0]{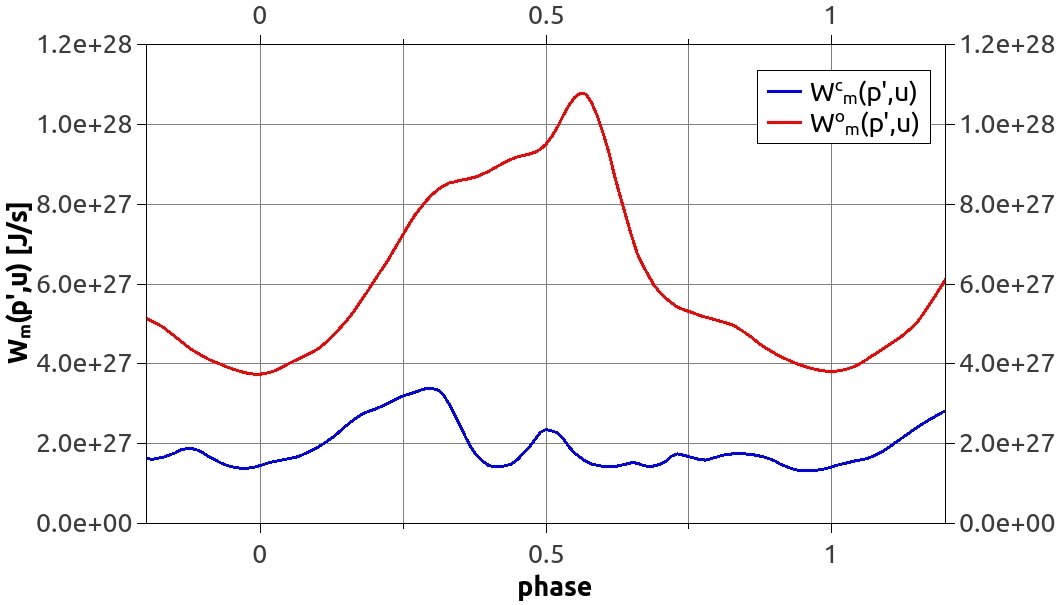}}
   \label{pic:mPu1} 
\subfigure[\sffamily]
   {\includegraphics[width=0.45\linewidth,angle=0]{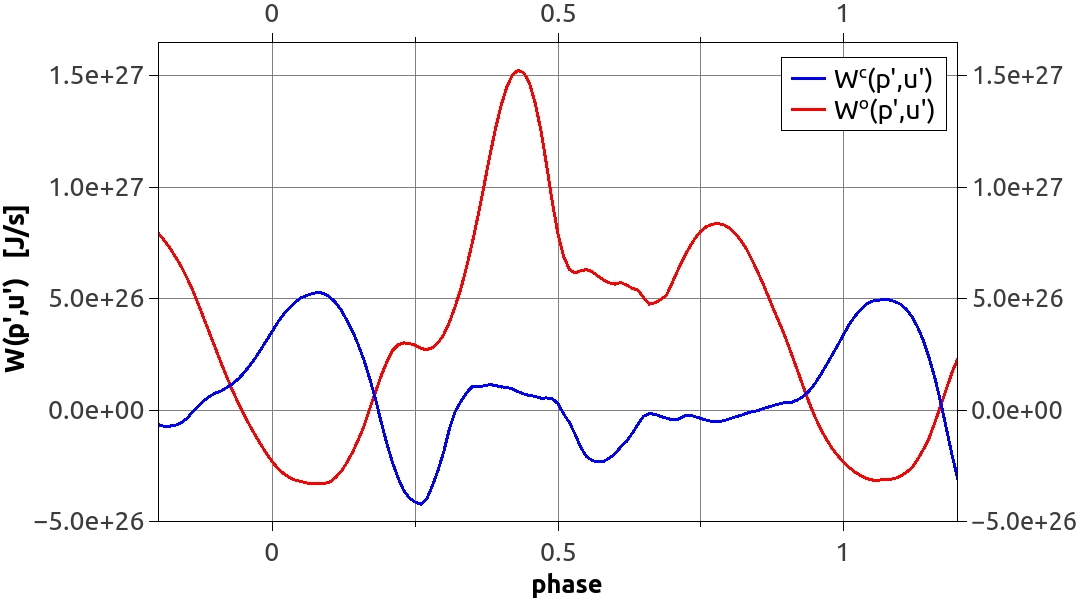}}
   \label{pic:P1u1}
   \hfill{}
\subfigure[\sffamily]
   {\includegraphics[width=0.45\linewidth,angle=0]{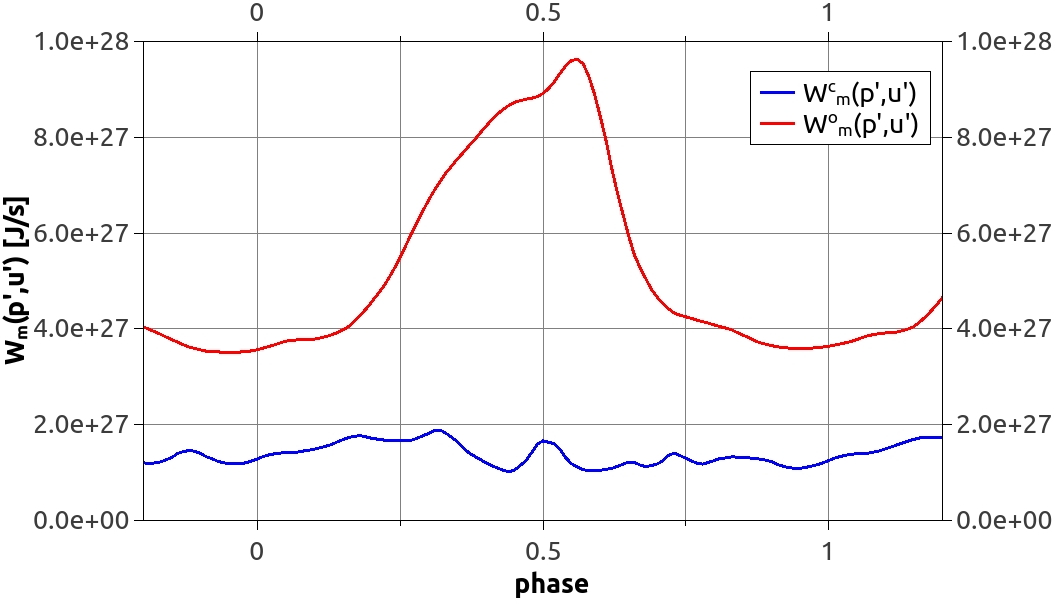}}
   \label{pic:mP1u1} 
\caption{The work integrals including the thermodynamic pressure for $\sim1$ period. The components used are indicated on the vertical axis of the different panels.  \label{fig:W_press}}
\end{figure*}

For the thermodynamic pressure $p$ the variation with phase is shown in Fig.~\ref{fig:W_press}. The work $W\left(p,u\right)$ can be seen to mainly follow the pulsation pattern for each zone, more clearly so in the overshooting zone.
The average values $\overline{W\left(p,\boldsymbol{u}\right)}$ (see Tab. \ref{tab:pw-values}) are positive everywhere, since excitation work can be observed at the top of the overshooting zone (see Fig. \ref{fig:W_press}~(a)).
In the convection zone the pdV-terms reach their smallest values during contraction while the average values are always positive and the modulus shows little variation indicating roughly constant activity. The panels on the right-hand side of Fig.~\ref{fig:W_press} show that in the overshooting zone
the modulus of the pdV-terms derived from fluctuation quantities reach their highest values right after maximal contraction at the beginning of the expansion. Cancellation effects in the work integral are most dramatic for the $W(p,\boldsymbol{u}^\prime)$ terms, reaching up to about three orders of magnitude (compare Fig~\ref{fig:W_press} (c) and (d)).

\begin{figure*}
\subfigure[\sffamily]
   {\includegraphics[width=0.45\linewidth,angle=0]{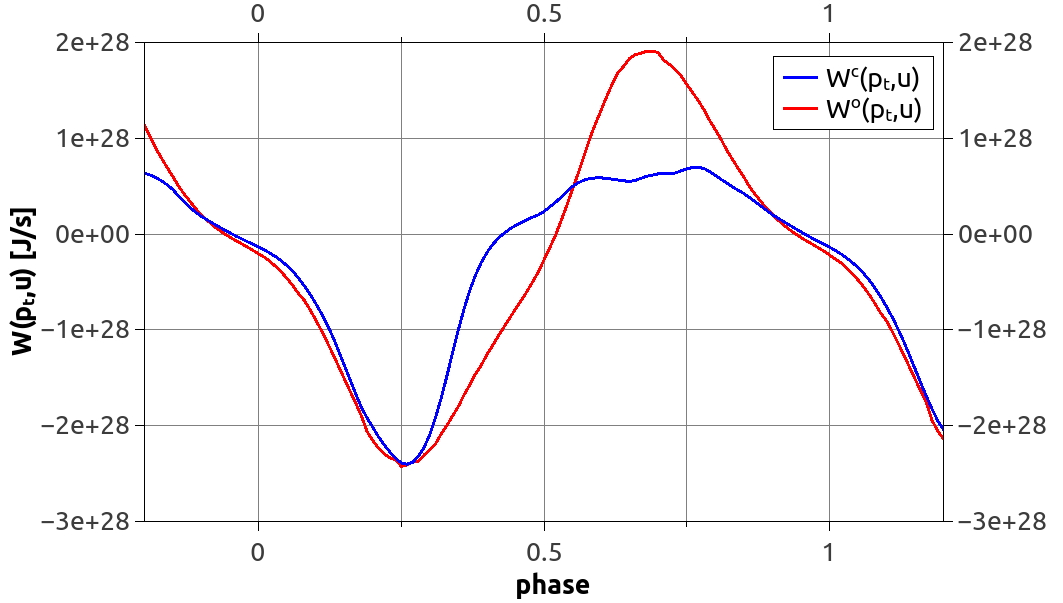}}
   \label{pic:Ptu}
   \hfill{}
\subfigure[\sffamily]
   {\includegraphics[width=0.45\linewidth,angle=0]{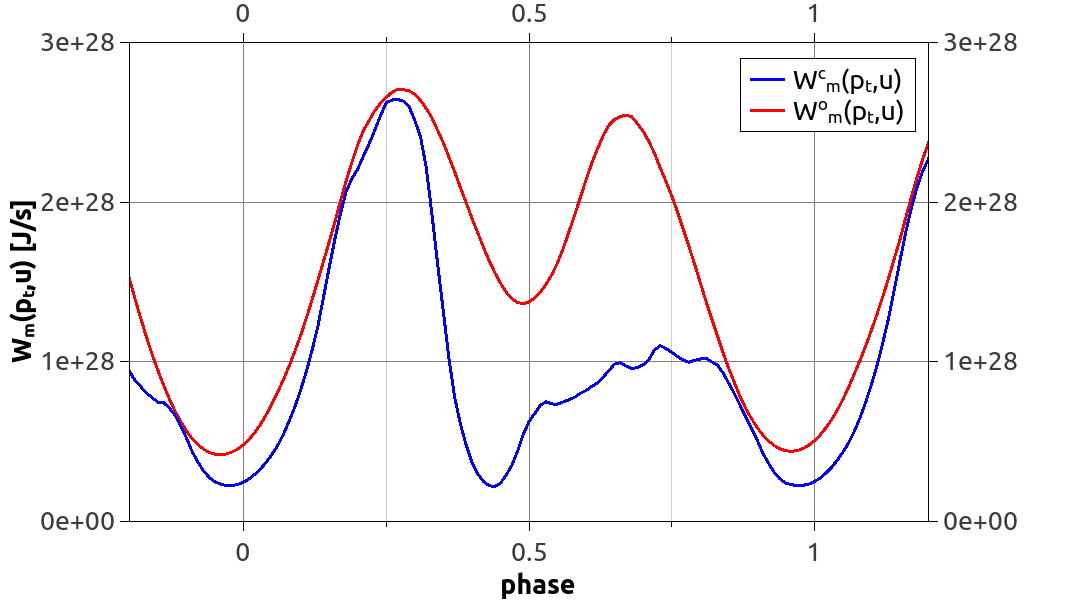}}
   \label{pic:mPtu}
\subfigure[\sffamily]
   {\includegraphics[width=0.45\linewidth,angle=0]{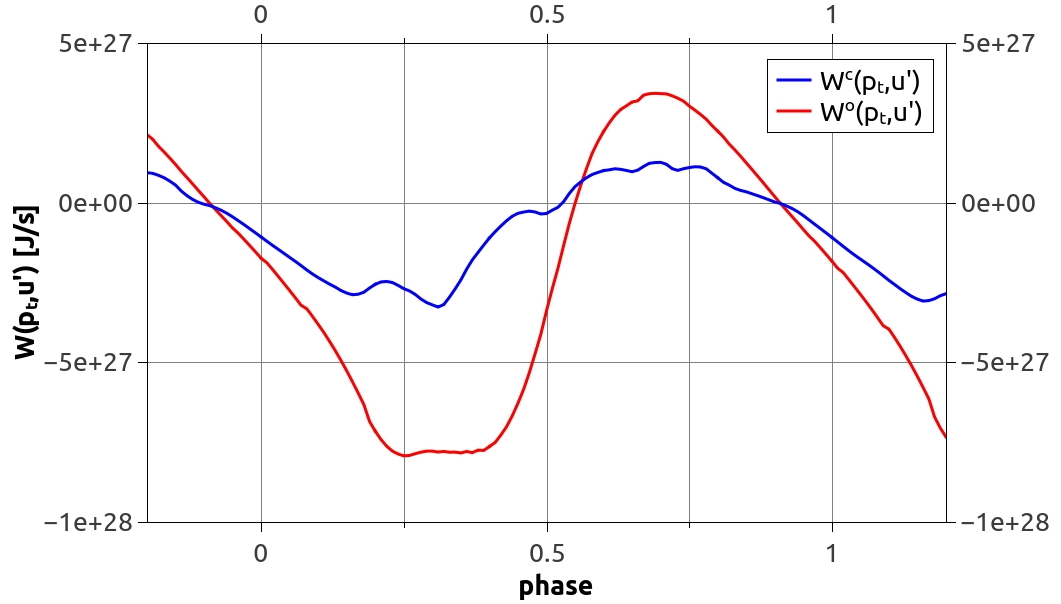}}
   \label{pic:Pt1u}
   \hfill{}
\subfigure[\sffamily]
   {\includegraphics[width=0.45\linewidth,angle=0]{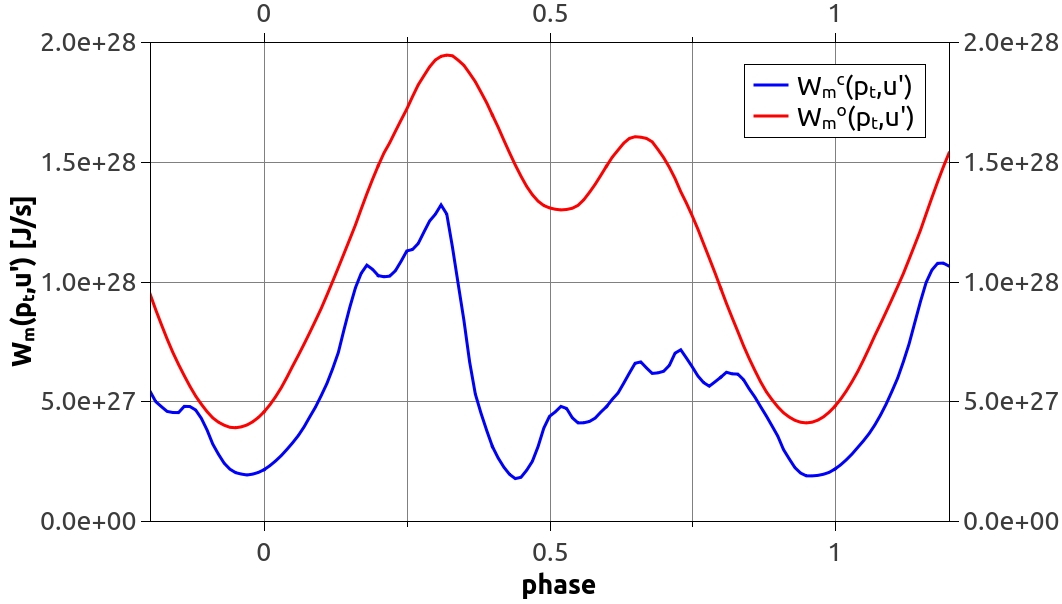}}
   \label{pic:mPt1u}
\subfigure[\sffamily]
   {\includegraphics[width=0.45\linewidth,angle=0]{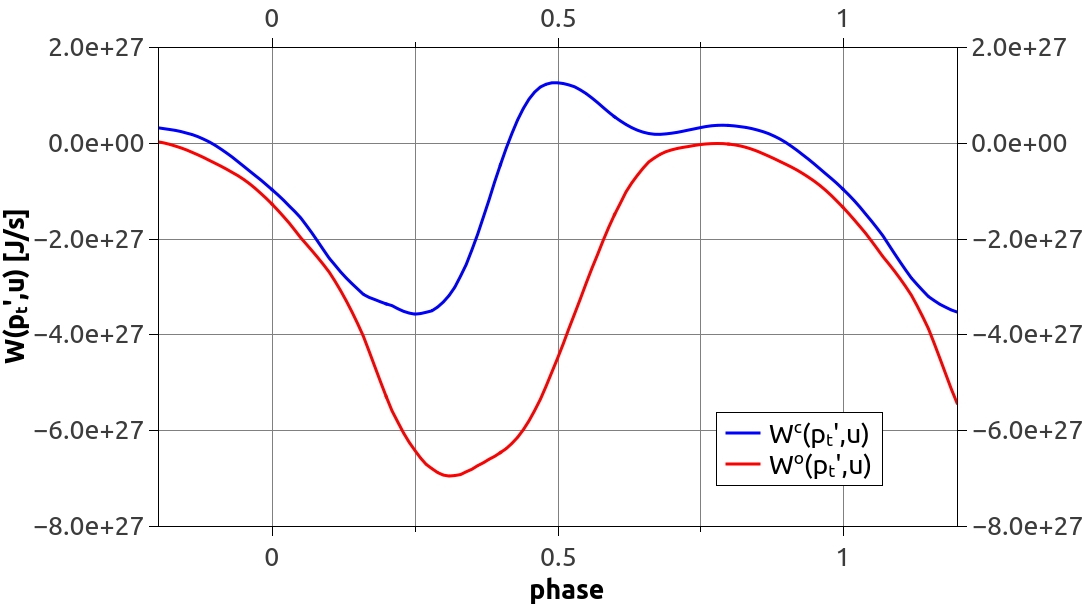}}
   \label{pic:Ptu1}
   \hfill{}
\subfigure[\sffamily]
   {\includegraphics[width=0.45\linewidth,angle=0]{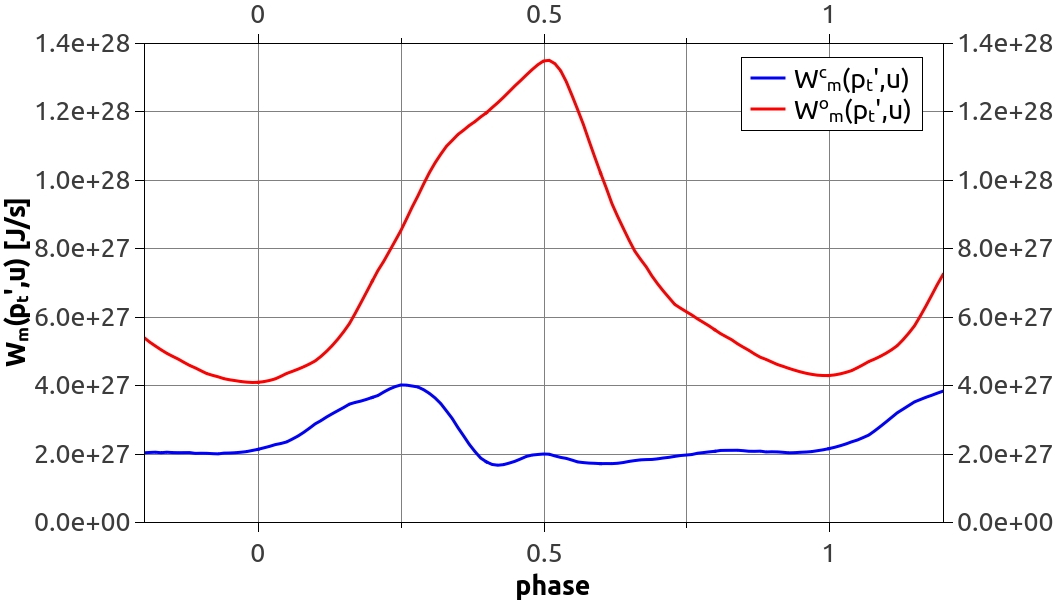}}
   \label{pic:mPtu1} 
\subfigure[\sffamily]
   {\includegraphics[width=0.45\linewidth,angle=0]{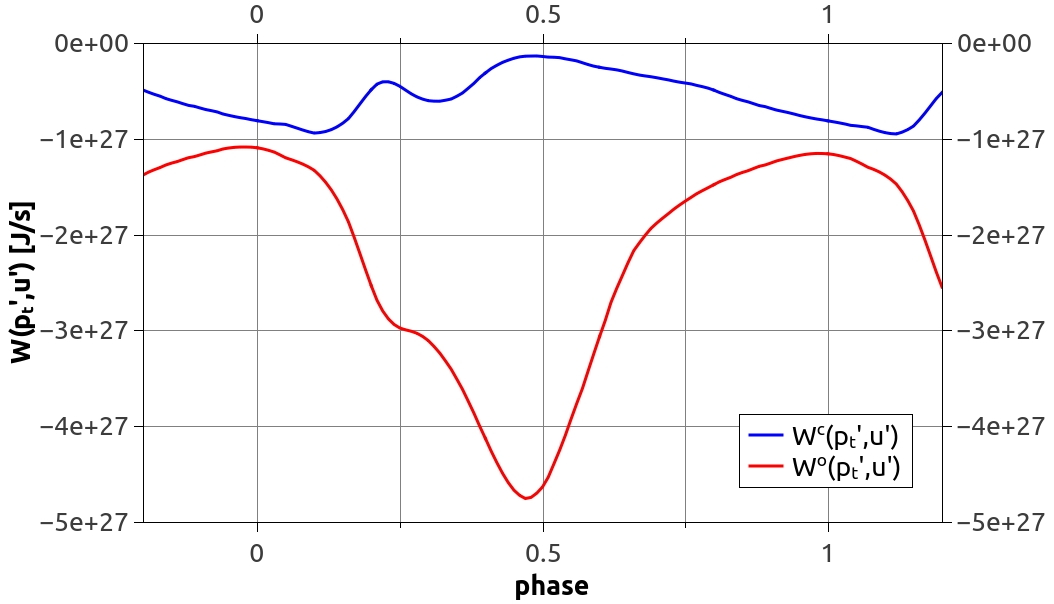}}
   \label{pic:Pt1u1}
   \hspace{1cm}
\subfigure[\sffamily]
   {\includegraphics[width=0.45\linewidth,angle=0]{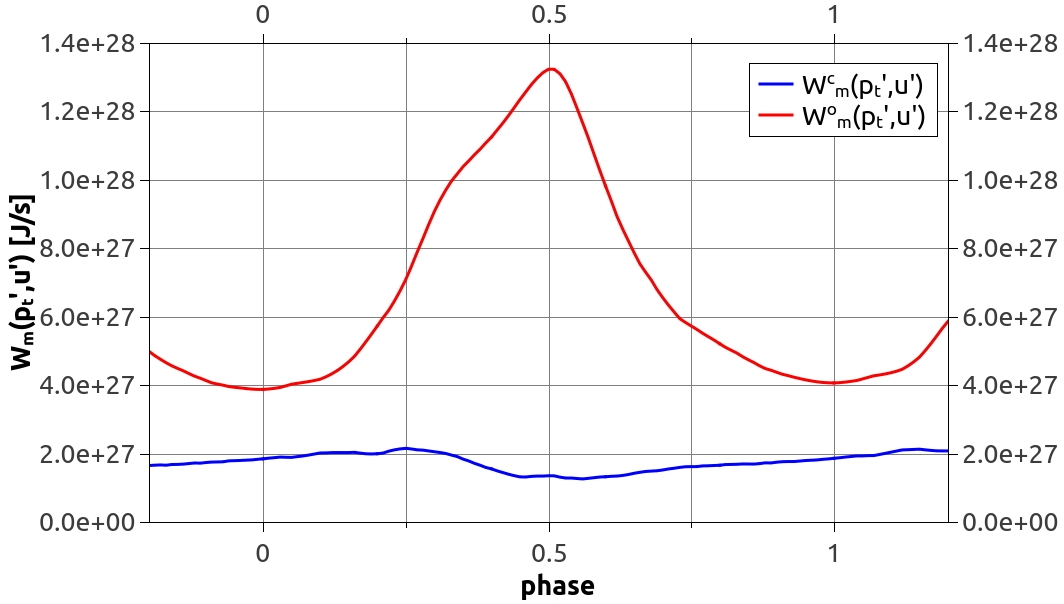}}
   \label{pic:mPt1u1} 
\caption{The work integrals including the turbulent pressure for $\sim1$ period. The components used are indicated on the vertical axis of the different panels.  \label{fig:W_pturb}}
\end{figure*}

For the turbulent pressure the pdV-terms follow the pulsation pattern if at least one component includes pulsation. In the case of $W\left(p_t^\prime,\boldsymbol{u}^\prime\right)$ one sees (Fig.~\ref{fig:W_pturb} (g) and (h)) that the dampening effects of the turbulent pressure perturbation are at full contraction at their maximum in the overshooting zone
but at their lowest in the convection zone. The modulus $W_m^c\left(p_t^\prime,\boldsymbol{u}^\prime\right)$ in the convection varies little.

 \section{Discussion and conclusions}   \label{sec:discussion}

While simulations of pulsating stars (Cepheids, RR Lyr) have been computationally one-dimensional for decades, assuming radial symmetry, multidimensional calculations start to be feasible only recently. They allow in particular the investigation of the pulsation-convection interaction which is considered a main open problem in the simulation of such stars (\cite{buch_rev09}). We have constructed a two-dimensional model regarding convection and pulsation in a Cepheid type star. The justification for adopting two dimensions at the present state of the matter can be taken from a comparison of 2D and 3D results in \cite{ger14}. These results show that (for models off RR Lyr stars) many quantities such as peak convective flux as a function of phase, growth rates or amplitudes are quite similar in 2D and 3D, differences being mainly visible in the slope of the light curve at various times in the descending branch. The aim of our investigation is to figure out basic properties of convection in the presence of pulsation and to draw inferences on existing ways of modelling these issues via time dependent convection models (TDCs) in a computationally 1D setting.

\subsection{Convection modelling in pulsating stars}  \label{subsec:TDC}
 
The TDCs just mentioned entail a number of coefficients not priorly known. With respect to the coefficient $\alpha_{\rm c} $, the calibration factor for obtaining the physically correct convective flux values from the TDC models of Kuhfu{\ss} and Stellingwerf, the following conclusions can be drawn from our work. 

In the He~{\sc ii} partial ionisation zone and its surrounding overshooting
zone convection is subject to large radiative losses, as is
witnessed by a low Peclet number ($Pe < 1$ predominantly). It is thus
difficult to calibrate $\alpha_{\rm c}$ separately from a coefficient 
$\alpha_{\rm r}$, connected to radiative losses, at least from our present model. 
After all, as explained in Sect.~3.1, these coefficients enter  essentially only in the combination $\alpha_{\rm c}\cdot\alpha_{\rm r}$ (or 
 $\alpha_{\rm c} / \alpha_{\rm r}$ in case of the Stellingwerf model) for our case of low Peclet numbers. 

Our models show that using one and the same coefficient $\alpha_{\rm c}$ for the convection zone (c.z.) and the overshoot zone (o.z.) would result in much too large overshooting predicted by the TDCs.  However,  $\alpha_{\rm c}$ and $\alpha_{\rm r}$ enter only in the combination described in Sec.~\ref{sec:flux_properties}. Consequently, including radiative losses by any choice of $\alpha_{\rm r}$ in the prescription for the convective flux cannot correct the discrepancy of too
large fluxes predicted by the model expressions for the overshooting
zone in comparison with the interior of the convection zone. At present,
the main effect of accounting for radiative losses, i.e. introducing
the factor $\eta$ in Eq.~(3), is to at least limit a more excessive
overshooting which would otherwise result from applying the prescription
of Kuhfu{\ss} (see Fig. 2). This model is more affected by that problem
than the model of Stellingwerf.

Accounting for the change of anisotropy of the contributions
to turbulent kinetic energy, however, is much more important.
To avoid excessive overshooting in the model requires introducing
a functional dependence in Eq.~(3) which can take care of
the fact that convective zones show stronger vertical fluxes
than horizontal ones while the opposite holds for overshooting
zones.

However, as we have shown, even after such corrections
a large discrepancy remains. This requires to at least
vary $\alpha_{\rm c,*}$ by a factor of 1.5 for the Stellingwerf
model and a factor of 4 for the Kuhfu{\ss} model to obtain
convective fluxes from the models which are consistent
with our numerical simulations. In addition, 
there is a clear systematic trend of $\alpha_{\rm c} $ as a function of depth within the convection zone,
see Fig.~\ref{fig:alpha_stripes}. The effect of this trend on modelling the pulsation in the traditional way is unknown.

Some contribution to this might actually be due to the coupling
between the mean structure and the turbulent kinetic energy
which is certainly more complex than expressed through Eq.~(3).

We should mention that the present comparison is already
the most favorable one for the TDC models in the sense that we only
take the model expressions to be meant to predict statistical averages:
variations for particular instants of phase for a specific
point in time may be larger and any attempt to interpret
the expressions in a point-wise sense is falsified immediately
by unacceptably large discrepancies (at least two orders of
magnitudes larger than the already large differences observed
for the averaged quantities considered here).

In this sense the present simulations unveil major
weaknesses of TDC models used to explain the pulsational
properties of Cepheids. The present simulations allow
to quantify the discrepancies expected from the TDC models
for a certain group of Cepheids. Since idealized simulations such
as those of \cite{gast_L11} come to similar conclusions
on a qualitative level, the discrepancies found are most
likely of a fundamental nature and thus have to be taken care
of. 

Otherwise one may fall into the trap of obtaining the right results for what might be the wrong reasons. Our results suggest that in 1D simulations one gets artificially large overshooting below the convection zone when applying identical values of $\alpha_{\rm c}$ in the c.z. and the o.z.. This  reinforces the doubts which have been expressed by \citet{smolec_mo08} in respect to the achievements on double mode pulsation in 1D Cepheid models. These doubts are based on the suspicion of too large overshooting in the models which have shown double mode pulsation in a broader parameter range than previously. 

One additional point is that 2D simulations tend to lead to larger overshooting than similar 3D simulations produce (\cite{muthsam95}). Considering this aspect as well, the problem of depth-dependency of  $\alpha_{\rm c}$ may well be even more severe than is borne out by the present investigation.

\subsection{Energetics of the convection zone} \label{subsec:ener}

Our results regarding the contribution of fluctuating quantities to the work integral reveal an intricate change of correlations as a function of phase. In Table~\ref{tab:pw-values} we compare values of the contributions to the work integral, integrated over the convection zone, and the overshoot zone. Considering contributions to the work integral where both entering quantities are fluctuating ones (instead of corresponding to the radial symmetric component) it turns out that no such combination can be disregarded: at least for one domain of integration every combination leads to a contribution of the same size as the others. That means that in principle each TDC model should faithfully represent the terms corresponding to all of the possible combinations (plus those with just one fluctuating quantity of course). It can be expected that work integral contributions with two fluctuating quantities will be more important in cases of stronger convection zones, since due to the more pronounced nonlinearity, their higher order terms will likely be larger.

A comparison of our simulations to TDC models regarding the energetics of the pulsation-convection interaction cannot be undertaken in the same manner as has been possible for the $\alpha$-factors discussed above. There, the relevant flux as delivered by the TDC models can, by a simple algebraic formula, directly be evaluated and the scaling factor be determined; the constancy or variability  of this factor then immediately leads to an assessment of the quality. This is to be contrasted to the case of the work integral. There, the relevant quantities in the TDC (turbulent energy or pressure plus furthermore others, not really provided by the usual TDCs) are not algebraic in nature. Instead, the equation for turbulent energy is an evolution equation where $e_t$ attains a value deemed physically correct only after it has been evolving for some time. Hence, comparison to the quantities from the multidimensional simulations cannot be performed on the basis of a temporal snapshot. At the same time it is not really possible to include an equation for the turbulent energy  $e_t$  in the multidimensional simulations: the simulations explicitly \textit{calculate} $e_t$ while  the equation for turbulent energy in the TDCs \textit{models} $e_t$. Consequently, there is a branching quite at the roots of the modelling processes between one- and multidimensional simulations which prevents a meaningful comparison of the two values of $e_t$ by solving the TDC equation alongside in multidimensional work. After all, there is no room to allow for backreaction of the TDC variant of $e_t$ in multidimensions. -- Our calculations however show that, for  modelling these values in the sense of TDCs, phase dependent properties of hydrodynamic quantities ought to be built into the models. After all, the contributions to the work integral (which are related to correlations of physical quantities) have a distinctive phase dependence which, moreover, is not in all cases similar in the convection and the overshoot zone. This cannot really come as a surprise given that convection in a pulsating star is not in a statistically steady state. Rather, it continually evolves in a specific way: at first convection cells form at the top of the convection zone at a certain phase, plumes then descend, overshoot into the stable zone below at some later phase and ultimately the remnants die off deep below the convection zone. Explicitly or implicitly, each TDC should account for such a sequence of events. 

\subsection{Requirements for time dependent convection models}

The simulations reported above suggest some guidelines for the development of revised TDCs to eventually be used in 1D simulations. Radiative losses have to be included in the model, but in addition the convective flux will have to evolve in an explicitly phase dependent manner. Therefore an additional evolution equation for $F_c$ seems inevitable, similar to the model of \cite{gough77}, further developed in \cite{balm92}, or the rarely used 3-equation variant of the Kuhfu{\ss} model. Moreover, the distribution of kinetic energy clearly changes as a function of depth in a non-trivial manner and most likely also features phase dependencies. Thus, a fully non-local Reynolds stress model as given by \cite{canuto93}, \cite{canuto97} and 
\cite{canuto99} may be required as a minimum level of complexity to substantially improve TDC models in comparison to those presently used.

\subsection{Further avenues of research} \label{subsec:outlook}

When finally turning our attention to further work along these lines, several items come to mind. One obvious goal consists in the construction of Cepheid models with a stronger He~{\sc ii} convection zone in order to more fully sample the parameter space. Such material may be crucial for improved TDCs for use by the 1D pulsation modelling community. 

One further goal is an investigation of the upper, H+He{\sc i}, convection zone and, hence, the atmosphere. We have already performed some initial runs with the necessary resolution which is considerably higher in the upper layers than in the models which we have reported here. Due to the fact that, with higher resolution, the H+He{\sc i} convection zone is much stronger than in our present models we have had some difficulty with the higher resolution models. In particular, throughout the atmosphere (and also below) there is vivid non-radial shock activity. This has lead to numerical problems which required the development of numerically suitable boundary conditions. Such conditions were devised in the context of solar granulation (\cite{grimm15}). A modification of these
conditions has recently been found to work for the Cepheid case. Shock activity with shocks up to a Mach number of about 4, triggered by convection, has also been found in preliminary calculations we have done with these new conditions. This highly dynamical situation in
the atmosphere of the Cepheid model also leads to a highly corrugated optical surface. To describe such a situation by a static or even a time-dependent model one-dimensional seems extremely challenging, and the implications for an analysis of photometric or spectroscopic data will need to be figured out.

A different avenue of research could center on investing pulsation in a ring (or, in 3D, a spherical shell) as initiated in preliminary work  (\cite{kup14}). Such simulations could shed light on the excitation of non-radial modes if they indeed show up, connected or not connected with convection, whatever the outcome of the simulations would be.

\section*{Acknowledgments}
This work has been supported by the Austrian Science Foundation, FWF grant P18224. We are thankful to
G. Houdek for supplying us with starting models. Calculations have been performed at the VSC 
cluster of the Vienna universities. FK and EM gratefully acknowledge support through the 
Austrian Science Foundation FWF, grant P21742. The work of FK has also been supported by FWF grant 25229.

\bibliographystyle{natbib}

\label{lastpage}

\end{document}